\documentclass[prd,nofootinbib,eqsecnum,final]{revtex4}
\usepackage{graphicx,color}
  \usepackage{bm}
   \usepackage{amsmath}
    \usepackage{amssymb}
     \usepackage{pifont}
      \usepackage{simplewick}
      \usepackage{stackrel}
\usepackage{tikz}
\usepackage[most]{tcolorbox}
\usepackage{rotating}

\newcommand{\nn}{\nonumber}

\newcommand{\const}{\text{const.}}

\newcommand{\ot}{\leftarrow}

\renewcommand{\(}{\left(}
\renewcommand{\)}{\right)}
\renewcommand{\[}{\left[}
\renewcommand{\]}{\right]}

\renewcommand{\vec}[1]{\bm{#1}}

\newcommand{\bn}{{\bar n}}

\newcommand{\bT}{{\vecb b_T}}

\newcommand{\vecb}[1]{\mbox{\boldmath $#1$}}
\newcommand{\vecbe}[1]{\mbox{\boldmath ${\scriptstyle #1}$}}

\newcommand{\sandwich}[3]{\left< #1 \right | #2 \left | #3 \right >}

\def\<{\langle}
\def\>{\rangle}

\def\d{\delta}

\def\m{\mu}

\def\t{\tau}

\def\({\left(}
\def\[{\left[}
\def\){\right)}
\def\]{\right]}

\def\ln{\hbox{ln}}

\def\le{ \left    }
\def\ri{ \right }

\usepackage[normalem]{ulem} 
\renewcommand\sout{\bgroup \color[rgb]{1,0,0} \ULdepth=-.5ex \ULset}

\begin{document}
\title{A short review on recent developments in TMD  factorization and implementation}

\author{Ignazio Scimemi}
\affiliation{Departamento de F\' isica Te\'orica and IPARCOS, \\ Universidad Complutense de Madrid,
Ciudad Universitaria, \\ 28040 Madrid, Spain}
\email{ignazios@ucm.es}

\begin{abstract}
In the latest years
the theoretical and phenomenological advances in the  factorization of  several collider processes  using the transverse momentum dependent distributions (TMD)
has   greatly increased.  I attempt here a short resume of the newest developments  discussing also the most recent perturbative QCD calculations. The  work is not strictly directed to experts in the field and   it wants to offer an overview of  the tools and concepts which are behind the TMD factorization and evolution.
  I consider both theoretical and phenomenological aspects, some of which have still to be fully explored.    It  is expected that  actual colliders and the  Electron Ion Collider (EIC) will provide important information in this respect.
\end{abstract}
\maketitle

\section{Introduction}

\label{sec:introduction}
The knowledge of the structure of hadrons is a leitmotiv for the study of quantum chromodynamics (QCD) for decades. Apart from the notions of quarks and gluons (we call them generically "partons" in the following), the  natural question is how the momenta of these particles are  distributed inside the hadrons and how the spin of hadrons is generated.
Phenomenologically it is possible to access  at this problem only in some particular  kinematical  conditions, as provided for instance in experiments  like (semi-inclusive) deep inelastic scattering, vector and scalar boson production, $\ell^+\ell^-\rightarrow$ hadrons or jets.
I review the basic principle which support this investigation.
 Let us consider, to start with, the cross section for di-lepton production in a typical Drell-Yan process $p p\rightarrow \ell^+\ell^-+X $ where $X$ includes all particles which are not directly measured. The cross section for this process can be written formally as
 \begin{align}
 \label{eq:1}
 \frac{d\sigma}{d Q^2}&\simeq\sum_{i,j=q,g}\int_0^1dx_1dx_2 {\cal H}_{ij}(Q^2,\mu^2) f_{i\leftarrow h}(x_1,\mu^2) f_{j\leftarrow h}(x_2,\mu^2)
 \end{align}
where  $Q^2$ is the virtual di-lepton invariant mass, $x_{i}$ are  the parton momenta fraction along a light-cone direction or Bjorken variables and  $f$ are the parton distribution functions (PDF). 
The r.h.s. of eq.~(\ref{eq:1})
assumes several notions  which, nowadays,  can be  found in textbooks.  In fact  a central hypothesis is a clear energy separation between the di-lepton invariant mass and the scale at which QCD cannot be treated perturbatively any more (we call it  the hadronization scale 
$\Lambda\sim {\cal O}(1)$ GeV), that is $Q^2\gg\Lambda^2$.
 Given this, one can factorize the cross section in a  perturbatively calculable part ${\cal H}$ and the  rest.  Formula~(\ref{eq:1}) represents  just a  first  term  of an "operator product expansion" of the cross section. The price to pay for this separation is the introduction of a factorization scale $\mu$
 which can be used   to resum logarithms in combination with renormalization group equations~\cite{Gribov:1972ri,Dokshitzer:1977sg,Altarelli:1977zs}.
  Another aspect,  which is  remarkable, is that  the non-perturbative part of the cross section can be also expressed as the product of two parton distribution functions. 
  This fact has two main  consequences: on the one hand,  all the non-perturbative  information of the process is included in the PDFs; on the other hand, the partons  belonging to different hadrons are completely disentangled. 
  In these conditions so the  longitudinal momenta of quarks and gluons can be reconstructed non-perturbatively and this fact has given rise to a large investigation  whose
  review  goes beyond the purpose  of this writing.
    
  The ideal description of  the process in eq.~(\ref{eq:1}) however becomes more involved in the case of more differential cross sections 
 ~\cite{Parisi:1979se,Collins:1981va,Collins:1984kg}.
  So, for instance, one can wonder whether a formula like
  \begin{align}
 \label{eq:2}
 \frac{d\sigma}{d Q^2 dq_T^2 dy}&\stackbin{?}{=}\sum_{i,j=q,g}\int d^2 \bT\; e^{-i \bT . {\vecb q_T}}\int_0^1dx_1dx_2 {\cal H}_{ij}(Q^2,\mu^2) F_{i\leftarrow h}(x_1,\bT,\mu^2) F_{j\leftarrow h}(x_2,\bT,\mu^2)
 \end{align}
  has any physical consistency\footnote{ I use the notation $\bT$ for 2-dimensional impact parameter, $-b_T^2={\bT^2}\geq 0$, $s$  is the center of mass energy of the process,
$$x_1=\frac{\sqrt{Q^2+q_T^2}}{\sqrt{s}}e^y
\quad
 x_2=\frac{\sqrt{Q^2+q_T^2}}{\sqrt{s}}e^{-y}.
$$
  }.
 The answer  to this question is necessarily more complex then in the case of eq.~(\ref{eq:1}) for the simple fact that a new kinematic scale, $q_T$, the transverse momentum of the di-lepton pair, has now appeared.  In this article I will concentrate on the description of the case 
 \begin{align}
 q_T\ll Q ,
 \label{eq:qTllQ}
 \end{align}
  which is interesting for a number of observables. The restriction to this kinematical regime represents also a limitation of the present approach which should be overcome with further  studies.  
 
 The study of factorization~\cite{Ji:2004wu,Becher:2010tm,Collins:2011zzd,GarciaEchevarria:2011rb,Chiu:2012ir,Echevarria:2012js}
  has lead finally to the conclusion that actually  eq.~(\ref{eq:2}) in not completely correct because the cross section for these kind of processes should instead be of the form 
  \begin{align}
 \label{eq:2p}
 \frac{d\sigma}{d Q^2 dq_T^2 dy}&{=}\sum_{i,j=q,g}\int d^2 {\bT}\; e^{-i \bT. {\vecb q_T}}\int_0^1dx_1dx_2 {\cal H}_{ij}(Q^2,\mu^2) F_{i\leftarrow h}(x_1,\bT,\zeta_1,\mu^2) F_{j\leftarrow h}(x_2,\bT,\zeta_2,\mu^2)
 \end{align}
  with $\zeta_1\zeta_2=Q^4$ and $\zeta_i$ being the rapidity scales.    Formula (\ref{eq:2p})  shows explicitly that the TMD  functions $F$  contain non-perturbative QCD information  different from the usual PDF, while they still allow   to complete disentangle QCD effects coming from different hadrons. These new nonperturbative QCD inputs can be written  in terms of  well defined matrix elements of field operators which can   be extracted from experiments or  evaluated  with  appropriate theoretical tools. These objectives
   require some discussion, which  I partially provide in this  text.
  
   The scale $\zeta$ is the authentic key stone of the  TMD  factorization.
   Its  origin is different from the  usual factorization scale $\mu$ and  because of this  it is allowed to perform a special resummation for this scale.
     This leads to the fact that  a consistent and efficient implementation of the $(\mu,\zeta)$ evolution is crucial for the prediction and extraction of TMDs from data. A possible implementation of the TMD evolution is   historically provided  by  Collins-Soper-Sterman (CSS)~\cite{Parisi:1979se,Collins:1981va,Collins:1984kg}.
    However a complete   discussion of more efficient alternatives has started more  recently~\cite{Aybat:2011zv,Echevarria:2012pw,DAlesio:2014mrz,Scimemi:2017etj,Scimemi:2018xaf}. The point is that the rapidity scale evolution  has both a perturbative and nonperturbative input, as it is actually provided by  (derivatives of) an operator matrix element (the so called soft-function). An efficient  implementation  and scale choice so  should  separate as much  as possible the  nonperturbative inputs  with different origin inside the cross-sections.
This target is not completely realized with the CSS implementation, while it can be achieved  with the  $\zeta$-prescription discussed  in the text.       
         This discussion is also relevant for multiple reasons. In fact various orders in perturbation theory are available already for unpolarized and polarized distribution and, in the  future,  one expects more results in this respect  for  many polarized distributions.
   When dealing with several perturbative orders,   the convergence of the perturbative series can be seriously undermined by an inappropriate choice of scales, and this is  a well known problem that can affect the theoretical error of any result. A more subtle issue  comes from the fact that the evolution corrections can also be of nonperturbative nature. It would be  certainly  clarifying a scheme in  which the nonperturbative effects of the evolution are clearly separated  
 from the instrinsic nonperturbative TMD effects.   Such a request  results to be important when several extraction of  TMD from data  are compared and also when a complete nonperturbative evaluation of TMD  can be provided.

 In the rest of this review I will try to give an idea   on how all these problems can be consistently treated, which can be useful also to explore  new and more efficient solutions.

\section{Factorization}
\label{sec:factorization}
\begin{figure*}[tb]
\centering
\includegraphics[width=\textwidth]{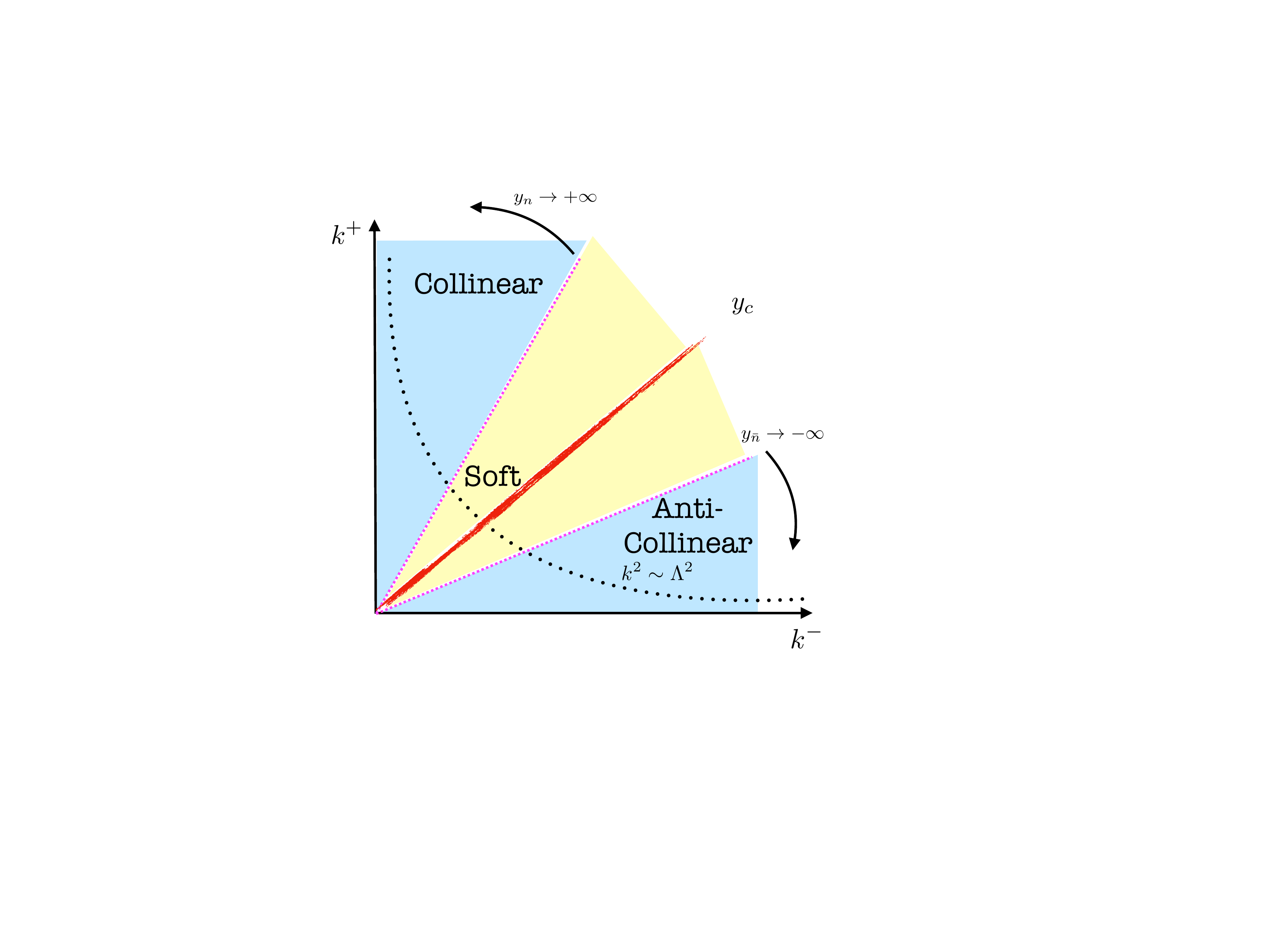}%
\vspace{-4cm}
\caption{Diagrams of regions for TMD factorization (orginal figure in \cite{Echevarria:2012js}).}
\label{fig:TMDjetRegimes}
\end{figure*}
The factorization of the cross sections into TMD matrix elements has been provided by several authors and it has been object of many discussions \cite{Parisi:1979se,Collins:1981va,Collins:1984kg,Ji:2004wu,Becher:2010tm,Collins:2011zzd,GarciaEchevarria:2011rb,Chiu:2012ir,Echevarria:2012js}.  We briefly review the main ideas here for the case of Drell-Yan. The process is  characterized  by  two initial hadrons which come from  opposite  collinear directions  and produce  two leptons in the final state plus unmeasured radiation. We identify collinear  (anti-collinear)
light-cone directions  $n$ ($\bar n$)  and $n^2=\bar n^2=0$, $n\cdot \bar n=1$ for  the momentum of colliding particles.   The momentum of collinear particles is $p=(p^+,p^-, p_\perp)$ with $n\cdot p=p^-,$ $\bar n\cdot p=p^+$ and $p_\perp =p-(n\cdot p) \bar n-(\bar n\cdot p)  n$ and $p^+\gg p_\perp\gg p_-$.  The  momenta of collinear particles are characterized  by the scaling $p\simeq Q (1,\lambda^2, \lambda)$ where $Q$ is  the di-lepton invariant mass  and $\lambda$ is a small parameter  $\lambda\sim \Lambda_{QCD}/Q$ being $\Lambda_{QCD}$ the hadronization scale. 
A reversed scaling of momentum is valid for anti-collinear particles, say $ p\simeq Q (\lambda^2,1, \lambda)$.
The soft radiation which entangles collinear and anti-collinear particles is homogeneous in momentum distribution (its momentum scales as $p\sim Q (\lambda,\lambda,\lambda)$)  and can be distinguished from the collinear radiation only for a different scaling of the components of the momenta.
Given this, it is  natural to divide the hadronic phase space  in regions as in fig.~\ref{fig:TMDjetRegimes}. In  this picture,  the collinear and  soft  regions are necessarily separated  by rapidity  and they all share the same energy $p^2\sim \Lambda^2$.

\subsection{Soft interactions and soft factor}
Because   the soft radiation is not finally measured, its interactions should be included  (and resummed) in the collinear parts, which become sensitive to  a rapidity scale which acts in a way  similar to the usual  factorization scale.
It is possible to define  the soft radiation through  a "soft factor", that is, by  an operator matrix element,
\begin{widetext}
\begin{align} \label{eq:softf}
S(\vecb k) &=
\int\frac{d^2\vecb b_T}{(2\pi)^2} e^{i\vecbe b _T \cdot \vecbe k}
\frac{{\rm Tr}_c}{N_c}
\sandwich{0}{\le[S_n^{T\dagger} \tilde S_\bn^T \ri](0^+,0^-,\vecb b_T)
\left[\tilde S^{T\dagger}_\bn S_n^T\right](0)}{0},
\end{align}
\end{widetext}
where  we have used the  Wilson line definitions~\cite{Idilbi:2010im,Idilbi:2010tc,GarciaEchevarria:2011md} appropriate for a Drell-Yan process,
\begin{align} \label{eq:SF_def2}
&S_{n}^T = T_{n(\bn)} S_{n}\,,
\quad\quad\quad\quad
\tilde S_{\bn}^T = \tilde T_{n(\bn)} \tilde S_{\bn}\,,
\nn\\
&S_n (x) = P \exp \left[i g \int_{-\infty}^0 ds\, n \cdot A (x+s n)\right]\,,
\nn \\
&T_{n} (x_T) = P \exp \left[i g \int_{-\infty}^0 d\tau\, \vec l_\perp \cdot \vec A_{\perp} (\infty^+,0^-,\vec x_T+\vec l_\perp \tau)\right]\,,
\nn \\
&T_{\bn} (x_T) = P \exp \left[i g \int_{-\infty}^0 d\tau\, \vec l_\perp \cdot \vec A_{\perp} (0^+,\infty^-,\vec x_T+\vec l_\perp \tau)\right]\,,
\nn\\
&\tilde S_\bn (x) = P\exp\le[-ig\int_{0}^{\infty} ds\, \bn \cdot A(x+\bn s) \ri]\,,
\nn\\
&\tilde T_{n} (x_T) = P\exp\le[-ig\int_{0}^{\infty} d\t\, \vec l_\perp \cdot \vec A_{\perp}(\infty^+,0^-,\vec x_T+\vec l_\perp\t) \ri]\,,
\nn\\
&\tilde T_{\bn} (x_T) = P\exp\le[-ig\int_{0}^{\infty} d\t\, \vec l_\perp \cdot \vec A_{\perp}(0^+,\infty^-,\vec x_T+\vec l_\perp\t) \ri]
\,.
\end{align}
The direct calculation of  the soft factor is all but  trivial and the  way the calculation is performed can influence directly the final formal definition of the transverse momentum dependent distribution used by different  authors. In fact a simple perturbative calculation shows that in the soft factor there are divergences which cannot be  regularized dimensionally (say, they are not explicitly ultraviolet (UV) or infrared (IR)) which occur when  the integration momenta are big  and aligned on the light cone directions. 
The divergences that arise in this configuration of momenta are generically called rapidity divergences and  regulated by a rapidity regulator.
One can understand the necessity of a specific regulator  observing that
the light-like Wilson lines  are invariant under the coordinate rescaling in their own light-like directions.
This invariance leads to an ambiguity in the definition of rapidity divergences.
Indeed, the boost of the collinear components of  momenta $k^+\to a k^+$, $k^-\to k^-/a$ (with $a$ an arbitrary number) leaves the soft function invariant, while in the limit $a\to\infty$  one obtains the rapidity divergent configuration.
Therefore  the soft function cannot be explicitly calculated without a regularization which breaks its boost invariance.
The coordinate space description of rapidity divergences, as well as, the counting rules for them have been derived in \cite{Vladimirov:2016qkd,Vladimirov:2017ksc}.
The nature of the divergences  in the soft factor has been studied explicitly in \cite{Echevarria:2013aca} at one loop and in \cite{Echevarria:2015byo} at NNLO, which conclude that, once all contributions are included, the  soft factor  depends only on ultraviolet and rapidity divergences (and  IR divergences are present only in the intermediate steps of the calculations, but not in the final result).
Different regulators have also shown to be more or less efficient  within different approaches to the calculations of transverse momentum dependent distributions. For instance  NNLO perturbative calculations for  unpolarized distributions, transversity and pretzelosity have been performed using de $\delta$-regulator  of \cite{Echevarria:2016scs,Gutierrez-Reyes:2017glx,Gutierrez-Reyes:2018iod} while for  the recent attempts of lattice calculations off-the-light-cone Wilson lines are preferred \cite{Hagler:2009mb,Musch:2010ka,Musch:2011er,Ji:2013dva,Ji:2014gla,Engelhardt:2015xja,Yoon:2016dyh,Yoon:2017qzo,Radyushkin:2017cyf,Orginos:2017kos,Ji:2018hvs}.
The discussion of the type of regulator involves usually another issue, which is also important for the complete definition of TMDs.
While collinear and soft sectors can be distinguished by rapidity, the choice of a rapidity regulator  forces a certain overlap of the two regions which should be removed, in  order to arrive to a consistent formulation of the factorized cross section. This is called "zero-bin" problem in Soft Collinear Effective Theory (SCET)~\cite{Manohar:2006nz}) and its solution is usually provided in any formulation of the factorization theorem.   The amount of the zero-bin overlap is usually fixed by the same soft function in some particular limit although it is generally impossible to define this subtraction in a unique (in the sense of regulator independent) form. 
Because of  this overlap one can find in the literature that the soft  function is used in a different way in different formulations of the factorization theorem. The evolution properties of TMDs however are independent of these subtleties and  they are the  same in all formulations. A possible rapidity renormalization scheme-dependance is traditionally fixed by requiring  $R^{-1}SR^{-1}=1$ (for this notation see discussion on sec.~\ref{subsec:TMDoperators}).

The factorization theorem to all orders in perturbation theory relies on the peculiar property of Soft function of being
at most linear in the logarithms generated by the rapidity divergences.
Then it comes natural  to factorize it  in two pieces \cite{Echevarria:2012js}, and in turn this feature  allows to define the individual TMDs.
Using the $\delta$-regulator  one can write to all orders in perturbation theory, as well as to all orders in the $\epsilon$-expansion (the UV divergences  are regulated  in dimensional regularization  $d=4-2\epsilon$)\cite{Echevarria:2015byo}.
\begin{align}\label{eq:splitting}&
\tilde{S}({\mathbf L}_\m,{\mathbf L}_{\sqrt{\d^+\d^-}}) =
\tilde{S}^\frac{1}{2}({\mathbf L}_\m,{\mathbf L}_{\d^+/\nu})\,
\tilde{S}^\frac{1}{2}({\mathbf L}_\m,{\mathbf L}_{\nu\d^-})
\,,
\end{align}
where tildes mark quantities calculated in coordinate space, $\nu$ is an arbitrary and positive real number that transforms as $p^+$ under boosts and we introduce the convenient notation
$${\mathbf L}_{X}\equiv\ln(X^2 \vecb \bT^2 e^{2\gamma_E}/4).$$
Despite the fact that the soft function is not measurable {\it per se},   its derivative provides the so called {\it rapidity anomalous dimension},
\begin{eqnarray}\label{eq:D_fromS}
{\cal D}=\frac{1}{2}\frac{d \ln \tilde S}{d\mathbf{l}_\delta}|_{\epsilon-\text{finite}}.
\end{eqnarray}
with $\mathbf{l}_\delta=\ln\(\mu^2/| \d_+\d_- |\)$. Because of its definition the rapidity  anomalous dimension ${\cal D}$  has both a perturbative (finite, calculable) part and a nonperturbative part.  This fact  should be  always taken into account despite the fact that many experimental data are actually marginally sensitive to the nonperturbative nature of the rapidity anomalous dimension.
A non-perturbative estimation of the evolution kernel with lattice has been recently proposed in \cite{Ebert:2018gzl} and I expect a deep discussion on this issue in the future.
A renormalon based calculation has also provided some approximate value for this nonperturbative contribution \cite{Scimemi:2016ffw}.

\subsection{TMD operators}
\label{subsec:TMDoperators}

Another fundamental ingredient in the formulation of the factorization theorem is represented  by the definition of the TMD operators that are involved. We use here the notation of \cite{Echevarria:2016scs}.
 The TMDs which appear in a Drell-Yan process  can be re-written   starting from the bare operators (here I consider only the quark case,  for simplicity)
\begin{align}\nn
O^{bare}_q(x,\vecb b_T)&=\frac{1}{2}\sum_X\int \frac{d\xi^-}{2\pi}e^{-ix p^+\xi^-}\left\{T\[\bar q_i \,\tilde W_n^T\]_{a}\(\frac{\xi}{2}\)
~|X\rangle \Gamma_{ij}\langle X|~\bar T\[\tilde W_n^{T\dagger}q_j\]_{a}\(-\frac{\xi}{2}\)\right\},
\label{def_PDF_op}
\end{align}
where $\xi=\{0^+,\xi^-,\vecb b_T\}$, $n$ and $\bn$ are light-cone vectors ($n^2=\bn^2=0,\; n\cdot\bn=1$), and $\Gamma$ is some Dirac matrix,
the repeated color indices $a$ ($a=1,\dots,N_c$ ) are summed up.
The representations of the color SU(3) generators inside the Wilson lines are the same as the representation of the corresponding partons.
The Wilson lines $\tilde W_n^T(x)$ are rooted at the coordinate $x$ and continue to the light-cone infinity along the vector $n$, where they are connected by a transverse link to the transverse infinity (that is indicated by the superscript $T$).
The \textit{bare}  or unsubtracted TMDs are  given then by the  hadronic matrix elements of
the corresponding \textit{bare} TMD operator:
\begin{eqnarray}\label{def:TMD_pdf_unsub}
\Phi_{f\ot N}(x,\vecb{b}_T)&=&\langle N|O^{bare}_f(x,\vecb{b}_T)|N \rangle.
\end{eqnarray}
These bare operators do not include for the moment  any  soft  radiation and  they are just collinear object (one can refer to them as "beam functions").  Because of boost invariance they can  be calculated in principle in any frame. However because of Wilson lines appearing  in their definition we have  to deal  with rapidity divergences and their regularization. The soft  interactions can be incorporated in the definition of the TMD  through an appropriate "rapidity renormalization factor" (which takes into account also a solution for the zero bin problem).  The  final form of the rapidity renormalization factor ($R$ in the following) is dictated by the factorization theorem.
The renormalized operators and the TMD are defined respectively as
\begin{eqnarray}\nn
O_{q}(x,\vecb b_T,\mu,\zeta)&=&Z_q(\zeta, \mu)R_q(\zeta,\mu)O^{bare}_{q}(x,\vecb b_T)\\
F_{f\ot N}(x,\vecb{b}_T;\mu,\zeta)&=&\langle N|O_f(x,\vecb{b}_T;\mu,\zeta)|N \rangle=Z_q(\zeta, \mu)R_q(\zeta,\mu)\Phi_{f\ot N}(x,\vecb{b}_T)
\label{eq:rz2}
\end{eqnarray}
and  $Z_q$  is the UV renormalization constant for TMD operators, and $R_q$  the rapidity renormalization factor. Both these factors are the same for particle and anti-particle however they are different  for quarks and gluons. These factors also occur in the same  way in parton distribution functions and fragmentation functions. The scales $\mu$ and $\zeta$ are the scales of UV and rapidity subtractions respectively.
 The factor $R_q$ is built out of the soft  factor and includes also the zero-bin corrections.
  There is a physical logic in this,  because the factor $R$ actually fixes how much soft radiation should be included inside a properly defined TMD.   In this respect  it is useful to specify in actual calculations how the  factor $R$ is derived.  For instance in \cite{Echevarria:2016scs}
  the authors first remove all rapidity divergences and perform
the zero-bin subtraction, and afterwards multiply by $Z$'s, and as a  result the $R$ factors depend both on rapidity and renormalization scales.

Different logic has been used by other authors. For instance,
 in \cite{Luebbert:2016itl}, the authors follow the ``Rapidity Renormalization Group'' introduced in \cite{Chiu:2011qc,Chiu:2012ir}, which is built in order to cancel the rapidity divergences through renormalization factors from the beam functions and soft factors independently although finally one achieves an equivalent resummation of  rapidity logarithms. In Ref.~\cite{Becher:2010tm,Gehrmann:2012ze,Gehrmann:2014yya} for TMDPDFs the soft  function is hidden in the product of two TMDs.

I conclude this section providing the actual definition of the rapidity renormalization factor $R$,

\begin{eqnarray}
R_f(\zeta,\mu)=\frac{\sqrt{S(\vecb{b}_T)}}{\textbf{Zb}},\quad f=q,g,
\end{eqnarray}
where $S(\vecb{b}_T)$ is the soft function and \textbf{Zb} denotes the zero-bin contribution, or in other words the soft overlap of the collinear and soft sectors which appear in the factorization theorem~\cite{Manohar:2006nz,Collins:2011zzd,GarciaEchevarria:2011rb,Echevarria:2012js,Echevarria:2014rua}.
Depending on the rapidity regularization, the zero-bin subtractions are related to a particular combination of the soft factors.
For instance the modified $\delta$-regularization \cite{Echevarria:2015byo} has been constructed such that the zero-bin subtraction is literally equal to the soft function: $\textbf{Zb}=S(\vecb{b}_T)$. The definition is non-trivial because it  implies a different regularized form for collinear Wilson lines $W_{n(\bn)}(x)$ and for soft Wilson lines $S_{n(\bn)}(x)$.
In the modified $\delta$-regularization, the expression for the rapidity renormalization factor is
\begin{eqnarray}\label{reg:R=1/S}
R_f(\zeta,\mu)\bigg|_{\delta\text{-reg.}}=\frac{1}{\sqrt{S(\vecb{b}_T;\zeta)}},
\end{eqnarray}
and this relation has been tested at NNLO in~\cite{Echevarria:2015usa,Echevarria:2015byo,Echevarria:2016scs}. We notice that due to the process independence of the soft function \cite{Collins:2011zzd,GarciaEchevarria:2011rb,Echevarria:2012js,Echevarria:2014rua,Collins:2004nx}, the factor $R_f$ is also
process independent. 
In the formulation of TMDs by Collins in \cite{Collins:2011zzd} the rapidity divergences are handled by tilting the Wilson lines off-the-light-cone. 
Then the contribution of the overlapping regions and soft factors can be recombined into individual TMDs by the proper combination of different soft functions with a partially removed regulator. 
This combination gives the factor $R^f$,
\begin{eqnarray}\label{reg:R=SS/S}
R^f(\zeta,\mu)\bigg|_{JCC}=\sqrt{ \frac{\tilde S(y_n,y_c)}{\tilde S(y_c,y_{\bar n}) \tilde S(y_n,y_{\bar n})}}.
\end{eqnarray}
The rest of logical steps remain the same as with the $\delta$-regulator.

An important aspect of factorization is  finally represented by the  cancellation of unphysical modes, the Glauber gluons. A check of this cancellation has been provided in
\cite{Collins:2011zzd,Gaunt:2014ska,Diehl:2015bca,Boer:2017hqr} and  I do not review it here.

\section{Matching at large $q_T$ (or small-$b$)}
\label{sec:factorization}
Once factorization is settled, the phenomenological analysis  of data using TMDs need more information  to  be practicable.
While  a complete nonperturbative calculation of TMD is not available at the moment  one can resort to  asymptotic limits of TMDs in order to achieve an approximate intuition of TMDs.
It turns out  that a valuable information can be achieved  in the limit of TMDs at  large transverse momentum. In this limit it is possible to "re-factorize" the TMDs in terms of Wilson coefficient and collinear parton distribution functions (PDF), following the usual rules for operator product expansion (OPE).
At operator level we have
\begin{eqnarray}\label{def_OPE}
O_f(x,\bT;\mu,\zeta)=\sum_{f'}C_{f\ot f'}(x,\bT;\mu,\zeta,\mu_b)\otimes O_{f'}(x,\mu_b)+\mathcal{O}\(\frac{\bT}{B_T}\),
\end{eqnarray}
where the symbol $\otimes$ is the Mellin convolution in variable $x$ or $z$ , and $f,\;  f'$ enumerate the  flavors of partons.
The running  on the scales $\mu$, $\mu_b$ and $\zeta$ is independent of the regularization scheme and it is dictated by the renormalization group equations that I will discuss in the next section.
Taking the hadron matrix elements of the operators we obtain the small-$b_T$ matching between the TMDs and their corresponding integrated functions,
\begin{eqnarray}
F_{f\ot N}(x,\bT;\mu,\zeta)=\sum_{f'}C_{f\ot f'}(x,\bT;\mu,\zeta,\mu_b)\otimes f_{f'\ot N}(x,\mu_b)
+\mathcal{O}\(\frac{\bT}{B_T}\),
\label{def_TMD_expansion}
\end{eqnarray}
The integrated functions (that is, the PDFs) depend only on the Bjorken variables ($x$ for PDFs) and the renormalization scale $\mu$,  while all the dependence on  the transverse coordinate $\vecb b_T$ and rapidity scale is contained in the matching coefficient and can be calculated perturbatively. The definition of the integrated PDFs is
\begin{align}\nn
f_{q\leftarrow N}(x)=\frac{1}{2}\sum_X\int \frac{d\xi^-}{2\pi}e^{-ix p^+\xi^-} \langle N| \left\{T\[\bar q_i \,\tilde
W_n^T\]_{a}\( \frac{\xi^-}{2} \) |X\rangle \gamma^+_{ij}\langle X|\bar T\[\tilde W_n^{T\dagger}q_j\]_{a}\(-\frac{\xi^-}{2} \) \right\} | N\rangle .
\end{align}

\begin{table}[t]
\begin{tabular}{l|c||c|c|c|c|c||}
\hline
 		& 			& Leading 	& Twist of 	& Maximum 			&  		& Mix\\
Name 	& Function 	& matching 	& leading 	& known order 		& Ref. 	& with\\
		& 			& function 	& matching 	& of coef.function	& 		& gluon
\\\hline\hline
unpolarized & $f_1(x,\vec b)$ & $f_1$ &tw-2 & NNLO ($a_s^2$) & \cite{Gehrmann:2014yya,Echevarria:2016scs} & yes
\\\hline
Sivers & $f_{1T}^\perp(x,\vec b)$ & $T$ & tw-3 & NLO ($a_s^1$) & \cite{Boer:2003cm,Ji:2006ub,Ji:2006vf,Koike:2007dg,Kang:2011mr,Sun:2013hua,Dai:2014ala,Scimemi:2018mmi,stv}*** &yes
\\\hline\hline
helicity & $g_{1L}(x,\vec b)$ & $g_1$ & tw-2 & NLO ($a_s^1$) & \cite{Bacchetta:2013pqa,Gutierrez-Reyes:2017glx,
Buffing:2017mqm,Scimemi:2018mmi} &yes
\\\hline
worm-gear T & $g_{1T}(x,\vec b)$ & $g_1$, $T$, $\Delta T$ & tw-2/3 & LO ($a_s^0$) &\cite{Kanazawa:2015ajw}* \cite{Scimemi:2018mmi} & yes
\\\hline
transversity & $h_1(x,\vec b)$ & $h_1$& tw-2 & NNLO($a_s^2$)  &  \cite{Gutierrez-Reyes:2018iod} & no
\\\hline
Boer-Mulders & $h_{1}^\perp(x,\vec b)$ & $\delta T_\epsilon$ & tw-3 & LO ($a_s^0$) & \cite{Scimemi:2018mmi}& no
\\\hline
worm-gear L & $h_{1L}^\perp(x,\vec b)$ & $h_1$, $\delta T_g$ & tw-2/3 & LO ($a_s^0$) &\cite{Kanazawa:2015ajw}* \cite{Scimemi:2018mmi}& no
\\\hline
pretzelosity** & $h_{1T}^\perp(x,\vec b)$ &  -- & tw-4 & --  & -- & --\\
\hline
\end{tabular}
\\
~
\\
*~~The calculation is done in the momentum space. The result is given for the moments of distribution.
\\
**~~ The pretzelosity can in principle be a twist-2 observable, however its twist-2 matching coefficient  has been found to be zero up to NNLO~\cite{Gutierrez-Reyes:2018iod}. Therefore one can conjecture that pretzelosity is  actually a twist-4 observable. Some arguments in favor of this can also be found in \cite{Chai:2018mwx}.
\\
*** The quark Sivers function  at NLO has a long story \cite{Ji:2006ub,Ji:2006vf,Koike:2007dg,Kang:2011mr,Sun:2013hua,Dai:2014ala}.
A complete  calculation is now available in~\cite{stv}.
\caption{\label{tab:final-table} Summary of available perturbative calculations of quark TMD distributions and their leading matching at small-b.}
\end{table}
In order to accomplish the calculation of the matching coefficients one uses eq.~(\ref{def_OPE}) on some particular states and solve the system for matching coefficients. For instance for twist-2 TMDs, since we are interested only in the leading term of the OPE, i.e. the term without transverse derivatives, it is enough to consider single parton matrix elements, with $p^2=0$.  The current status of these calculations for  quark  distributions is resumed in 
tab.~\ref{tab:final-table}. Less information is generally available in the case of gluon TMDs.   Basically  the matching coefficients for  unpolarized gluons are known at NNLO \cite{Scimemi:2018mmi}  and linearly polarized gluons at NLO \cite{Gutierrez-Reyes:2017glx}.
In general the  TMDs which match onto collinear  twist-3 functions are much less known, which reflects the difficulty of the computations.
It would be very useful to have a better knowledge of all these less known functions at higher perturbative order before the advent of Electron Ion Collider (EIC).  In the rest of this section I focus on unpolarized quark  distributions which offer also an important understanding on the power of the TMD factorization. 
The  necessity of  a complete NLO estimation of all TMDs is both theoretical and phenomenological. Actually
a difficulty of the TMD  extraction from data  is due to the fact that it is  a nontrivial function of two variables (Bjorken $x$ and transverse momentum) so that  a complete mapping on a plane is necessary.  This target is achievable thanks to the factorization of the cross section and the consequent extraction of the TMD evolution part, which is process independent.
A  second  important information comes from the asymptotic limit of  the TMD,  which is perturbatively calculable.  The simple LO expressions for the TMD in general do not provide much information (they are  just constants), so that in order to achieve a wise modeling a  NLO calculation is always necessary.   The  higher order calculations  allow  also to test the 
stability with respect to the scales that match the TMD perturbative and nonperturbative parts. For  the unpolarized case a study in this sense can be found in \cite{Scimemi:2017etj} both  for high energy and low energy data. Using a LO calculations  one cannot even quantify this error. 
Finally, another lesson that comes  from the analysis of the unpolarized case is that  a good portion of the TMD is tractable starting from their asymptotic expansion for large transverse momenta.
 In any case even a  10$\%$ average precision of the SIDIS cross section at EIC will need  a NLO theoretical  input.

\section{Evolution}
\label{sec:factorization}

\begin{table}[h]
\renewcommand{\arraystretch}{1.5}
\begin{tabular}{l|c|| c | c | c | c |  c|}
\hline
 & ~~\parbox[b]{1.7cm}{rapidity evolution scale}~~ & ~~\parbox[b]{1.7cm}{TMD anomalous dimension}~~ & ~~\parbox[b]{1.7cm}{cusp anomalous dimension}~~ &  ~~\parbox[b]{1.7cm}{vector form factor anomalous dimensions}~~ & \parbox[b]{1.7cm}{rapidity anomalous dimension} 
 \\ \hline\hline
 \cite{Echevarria:2012pw,Echevarria:2015usa,Scimemi:2017etj,Scimemi:2018xaf} & $\zeta$ & $\gamma_F$ & $\Gamma$ & $\gamma_V$ &$\mathcal{D}$ 
\\\hline
\cite{Collins:2011zzd,Aybat:2011zv} & $\zeta$ & $\gamma_F ~(=\gamma_D)$& $\frac{1}{2}\gamma_K$ & ~~$-\gamma_F(g(\mu);1)$~~ & $-\frac{1}{2}\tilde K$
\\\hline
\cite{Becher:2010tm,Becher:2011xn,Gehrmann:2014yya} & -- & -- & $\Gamma_{cusp}$ & $2\gamma^q$ & $\frac{1}{2}F_{f\bar f}$
\\\hline
\cite{Chiu:2012ir} & $\nu^2$ & $\gamma_\mu^{f_\perp}$ & $\Gamma_{cusp}$ & -- & $-\frac{1}{2}\gamma_\nu^{f_\perp}$
\\
\hline
\end{tabular}
\caption{\label{tab:ADs} Notation for TMD anomalous dimensions used in the literature.}
\end{table}

The  factorization scale dependence of the  TMDs can be established  starting from their defining operators and from eq.~(\ref{eq:rz2}),
\begin{eqnarray}
\mu^2 \frac{d}{d\mu^2}O_f(x,\vecb b_T)=\frac{1}{2}\gamma^f(\mu,\zeta)O_f(x,\vecb b_T) &\quad\hookrightarrow\quad&
\mu^2 \frac{d}{d\mu^2} F_{f\ot h}(x,\bT;\mu,\zeta)=\frac{\gamma^f_F(\mu,\zeta)}{2}F_{f\ot h}(x,\bT;\mu,\zeta),
\label{RGE:mu}
\end{eqnarray}
 in an usual way.
 The equation (\ref{RGE:mu}) is a standard renormalization group equation (which comes from the renormalization of the ultraviolet divergences),  the function $\gamma_F(\mu,\zeta)$ is called the TMD anomalous dimension and it contains both single and double logarithms.
  The same eq.~(\ref{eq:rz2}) can be used  to write the running  with respect to the rapidity scale, $\zeta$,  which is fixed from the knowledge of soft interactions (see discussion in \cite{Echevarria:2015byo}, also in \cite{Chiu:2011qc})
  and comes from the factorization of rapidity divergences (see e.g. ~\cite{Echevarria:2015usa,Vladimirov:2016qkd,Vladimirov:2017ksc}). Given that the soft factor
  is the same for initial and final states, the rapidity scale evolution is universally valid for TMD parton distribution functions and TMD fragmentation functions,  and it is also spin-independent (so it is the same also for TMDs at higher twist),
\begin{eqnarray}\label{RGE:zetaO}
\zeta \frac{d}{d\zeta}O_f(x,\bT)=-\mathcal{D}^f(\mu,\vecb{b}_T)O_f(x,\bT).
&\quad\hookrightarrow\quad&
\zeta\frac{d}{d\zeta}F_{f \ot h}(x,\bT;\mu,\zeta)= -\mathcal{D}^f(\mu,\bT)F_{f\ot h}(x,\bT;\mu,\zeta),
\end{eqnarray}
  The function $\mathcal{D}(\mu,\bT)$ is called the rapidity anomalous dimension and actually  one has  $\mathcal{D}(\mu,\bT)\equiv\mathcal{D}(\mu,|\bT|)$. 
  Several notations for
  rapidity anomalous dimensions have been used in the literature. The notations $\gamma_F$ and $\mathcal{D}$, used in this article, were suggested in~\cite{Echevarria:2012pw}. For convenience we list some popular notations and their relation to our notation in the table~\ref{tab:ADs}.

One has a different anomalous dimension for quarks and gluons, and the QCD properties of  exponentiation  implies the so-called Casimir scaling of anomalous dimension $\mathcal{D}$, see \cite{Echevarria:2015byo},
\begin{eqnarray}
\frac{\mathcal{D}^q}{\mathcal{D}^g}=\frac{C_F}{C_A}=\frac{N_c^2-1}{2N_c^2},
\end{eqnarray}
which has been checked up to three loops \cite{Li:2016ctv,Vladimirov:2016dll}.

The consistency of the differential equations (\ref{RGE:mu}-\ref{RGE:zetaO}) implies that the cross-derivatives of the anomalous dimension are equal to each other (\cite{Echevarria:2015byo,Chiu:2011qc}),
\begin{eqnarray}
\mu^2 \frac{d}{d\mu^2}\(-\mathcal{D}^f(\mu^2,\vecb{b}_T)\)=\zeta\frac{d}{d\zeta}\(\frac{\gamma^f(\mu,\zeta)}{2}\)=-\frac{\Gamma_{cusp}^f}{2}.
\label{eq:cusp1}
\end{eqnarray}
 From Eq.~(\ref{eq:cusp1}) one finds that the anomalous dimension $\gamma$ is
\begin{eqnarray}
\gamma^f=\Gamma_{cusp}^f\mathbf{l}_\zeta-\gamma_V^f,
\end{eqnarray}
where we introduce the notation
\begin{eqnarray}\label{def_logarithms}
\mathbf{l}_X\equiv \ln\(\frac{\mu^2}{X}\).
\end{eqnarray}
 The large-$q_T$ expansion of the TMD introduces also another  evolution scale, which is needed for the matching Wilson coefficients,
that  can be obtained by deriving both sides of eq.~(\ref{def_OPE}).
In the case of the unpolarized  TMDs this is provided by the  DGLAP\footnote{DGLAP is an acronym for Dokshitzer, Gribov, Lipatov, Altarelli, Parisi~\cite{Gribov:1972ri,Dokshitzer:1977sg,Altarelli:1977zs}.} equations
\begin{align}
\mu_b^2 \frac{d}{d\mu_b^2}O_f(x,\mu_b)=\sum_{f'}P_{f\ot f'}(x)O_{f'}(x,\mu_b),
\end{align}
where $P$ are the DGLAP kernels for the PDF. Similar equations hold for unpolarized TMD fragmentation functions (at NLO  one can check \cite{Moch:1999eb,Mitov:2006wy}).
It is useful to recall also the running of the matching coefficient with respect to the rapidity scale (we set $\mu_b=\mu$)
\begin{eqnarray}
\zeta \frac{d}{d\zeta}C_{f\ot f'}(x,\vecb{b}_T;\mu,\zeta)=-\mathcal{D}^f(\mu,\vecb{b}_T) C_{f\ot f'}(x,\vecb{b}_T;\mu,\zeta),
\end{eqnarray}
The solutions of these differential equations  are
\begin{eqnarray}
C_{f\ot f'}(x,\vecb{b}_T;\mu,\zeta)&=& \exp\(-\mathcal{D}^f(\mu,\vecb{b}_T)\mathbf{L}_{\sqrt{\zeta}}\)
\hat C_{f\ot f'}(x,\mathbf{L}_\mu)\ .
\end{eqnarray}
This defines the reduced matching coefficients $\hat C$ whose renormalization group evolution  equations are
\begin{eqnarray}
\mu^2 \frac{d}{d\mu^2}\hat C_{f\ot f'}(x,\mathbf{L}_\mu)=\sum_r \hat C_{f\ot r}(x,\mathbf{L}_\mu)\otimes K_{r\ot f'}^f(x,\mathbf{L}_\mu),
\end{eqnarray}
with the kernel $K$ 
\begin{eqnarray}
K^f_{r\ot f'}(x,\mathbf{L}_\mu)&=&\frac{\delta_{rf'}}{2}\(\Gamma_{cusp}^f\mathbf{L}_\mu-\gamma_V^f\)-P_{r\ot f'}(x).
\end{eqnarray}
Using these equations one can find the  expression for the logarithmical part of the matching coefficients at any given order, in terms of the anomalous dimensions and the finite part of the coefficient at one order lower.
It is convenient to introduce the notation for the $n$-th perturbative order:
\begin{align}
\label{eq:Clognotation}
\hat C^{[n]}_{f\ot f'}(x,\mathbf{L}_\mu)=\sum_{k=0}^{2n}C_{f\ot f'}^{(n;k)}(x)\mathbf{L}_\mu^k\ .
\end{align}
Given the knowledge of the coefficient at order $n-1$ one can reconstruct 
all the terms  with $k\neq 0$  at order $n$ in this series. So  finally any higher order calculation provides new informations on terms  $C_{f\ot f'}^{(n;0)}$. A resume of the present status of available calculations is provided in tab.~\ref{tab:final-table} .

\section{Implementation of  TMD formalism and TMD extraction from data}

The implementation of  TMD formalism  and its phenomenological application is not trivial and eq.~(\ref{eq:2p})  should be written more carefully in order to describe correctly each single experiment.
As an example let me review the case of the study of unpolarized TMD parton distribution functions in Drell-Yan and Z-boson production following \cite{Scimemi:2018xaf}.

Namely I consider the process   $h_1+h_2\to G(\to ll')+X$, where $G$ is the electroweak neutral gauge boson, $\gamma^*$ or $Z$. The incoming hadrons $h_i$ have momenta $p_1$ and $p_2$ with $(p_1+p_2)^2=s$. The gauge boson decays to the lepton pair with momenta $k_1$ and $k_2$. The momentum of the gauge boson or equivalently the invariant mass of lepton pair is $Q^2=q^2=(k_1+k_2)^2$. The differential cross-section for the Drell-Yan process can be written in the form \cite{Drell:1970wh,Altarelli:1978id}
\begin{eqnarray}\label{th:ds1}
d\sigma=\frac{d^4q}{2s}\sum_{G,G'=\gamma,Z}L_{GG'}^{\mu\nu}W_{\mu\nu}^{GG'}\Delta_G(q)\Delta_{G'}(q),
\end{eqnarray}
where $1/2s$ is the flux factor, $\Delta_G$ is the (Feynman) propagator for the gauge boson $G$. The hadron and lepton tensors are respectively
\begin{eqnarray}
W^{GG'}_{\mu\nu}&=&\int \frac{d^4 z}{(2\pi)^4} e^{-iqz}
\langle h_1(p_1)h_2(p_2)|J^G_\mu(z)J^{G'}_\nu(0)|h_1(p_1)h_2(p_2)\rangle,
\\
\label{th:leptonTensor}
L_{\mu\nu}^{GG'}&=&\int\frac{d^3k_1}{(2\pi)^3 2E_1}\frac{d^3k_2}{(2\pi)^3 2E_2} (2\pi)^4 \delta^4(k_1+k_2-q)
 \langle l_1(k_1) l_2(k_2)|J_{\nu}^G(0)|0\rangle\langle 0|J_{\mu}^{G'}(0)|l_1(k_1) l_2(k_2)\rangle,
\nn \\ &&
\end{eqnarray}
where $J_\mu^G$ is the electroweak current.
Within the TMD factorization, one obtains the following expression for the unpolarized hadron tensor (see e.g.~\cite{Tangerman:1994eh})
\begin{eqnarray}\label{th:hadron_tensor}
W_{\mu\nu}^{GG'}&=&\frac{-g_{T\mu\nu}}{\pi N_c}|C_V(q_T,\mu)|^2\sum_{f,f'}z_{ff'}^{GG'} \int \frac{d^2\vec b_T}{4\pi}e^{i({\bT} . {\vecb q_T} )}
F_{f\ot h_1}(x_1,\vec b_T;\mu,\zeta_1)F_{f'\ot h_2}(x_2,\vec b_T;\mu,\zeta_2)+Y_{\mu\nu},
\end{eqnarray}
where $g_T$ is the transverse part of the metric tensor and the summation runs over the active quark flavors. The variable $\mu$ is the hard factorization scale.  The variables $\zeta_{1,2}$ are the scales of soft-gluons factorization, and they fulfill the  relation $\zeta_1\zeta_2\simeq Q^4$. In the following, we consider the symmetric point $\zeta_1=\zeta_2=\zeta=Q^2$. 
The factors $z_{ff'}^{GG'}$ are the electro-weak charges and they are given explicitly in \cite{Scimemi:2018xaf}. 
The factor $C_V$ is the matching coefficient of the QCD neutral current to the same current  expressed in terms of collinear quark fields. The explicit expressions for $C_V$ can be found in \cite{Kramer:1986sg,Matsuura:1988sm,Idilbi:2006dg}.

 Finally, the term $Y$ denotes the power corrections to the TMD factorization theorem (to be distinguished from the power corrections to the TMD operator product expansion). The $Y$-term is of order $q_T/Q$ and is composed of TMD distributions of  higher dynamical twist and in principle it can also include factorization breaking  terms.  These contributions appear each time the condition in eq.~(\ref{eq:qTllQ}) is broken. It is a subtle issue to quantify exactly the magnitude of  the ratio $q_T/Q$ where the $Y$-terms become important.
  A phenomenological study in \cite{Scimemi:2018xaf} and a more formal study in the large-$N_c$ limit (that is, the limit of large number of colors) in \cite{Balitsky:2017gis} have found  a reasonable upper value  $(q_T/Q)_{max}\sim 0.2$.  A study which takes into account the structure of operators in the type of corrections  has been started in \cite{Ebert:2018gsn}.

  In general the $Y$-terms  should be included  when the di-lepton invariant mass is of order a few GeV (this is the case for instance of  HERMES experiment and, perhaps to a possibly less extent, COMPASS) or when the experimental precision is extreme (as it possibly happens with ATLAS experiment).
This is issue is important phenomenologically and involves the study of  cross sections  with the inclusion of factorization breaking contributions.
Some  recent suggestion have appeared in \cite{Collins:2016hqq,Gamberg:2017jha}  which have still to be tested  phenomenologically.  One should remark however that  the implementation of these  factorization breaking correction  strongly  depends on the  fact that  the factorized part of the cross section is correctly realized and phenomenologically tested.  More studies on this issue are necessary in the future.

Evaluating the lepton tensor, and combining together all factors one obtains the cross-section for the unpolarized Drell-Yan process at  leading order of  TMD factorization, in the form \cite{Collins:1984kg,Davies:1984sp,Ellis:1997ii,Becher:2010tm,GarciaEchevarria:2011rb,Collins:2011zzd}
\begin{eqnarray}\label{th:Xsec_gen}\nn
\frac{d\sigma}{dQ^2dyd(q_T^2)}&=& \frac{4\pi}{3N_c}\frac{\mathcal{P}}{sQ^2}
\sum_{GG'}z_{ll'}^{GG'}(Q)\sum_{ff'}z_{ff'}^{GG'} |C_V(Q,\mu)|^2
\int \frac{d^2\vec b_T}{4\pi}e^{i(\vec b_T\vec q_T)}
F_{f\ot h_1}(x_1,\vec b_T;\mu,\zeta)
F_{f'\ot h_2}(x_2,\vec b_T;\mu,\zeta)+Y,
\end{eqnarray}
where $y$ is the rapidity of the produced gauge boson.  The factor $\mathcal{P}$ is a part of the lepton tensor and contains information on the fiducial cuts. This factor provides important information  on the actual measured leptons and should be always included  when the relative experimental information is provided.

The evaluation of this cross section requires a correct implementation also of the  evolution and perturbative information of the TMDs.
In the rest of this section I dedicate particular  emphasis to  the evolution parts making  the point that passing from the all-order formal knowledge of the  factorized cross-section
 to the finite-order practical  usage requires the discussion of  some subtle points.

\subsection{The treatment of TMD evolution}

The TMD evolution is resumed by the following equations
\begin{eqnarray}\label{def:TMD_ev_UV}\nn
\mu^2 \frac{d}{d\mu^2} F_{f\ot h}(x,\bT;\mu,\zeta)&=&\frac{\gamma_F(\mu,\zeta)}{2}F_{f\ot h}(x,\bT;\mu,\zeta)
\quad \text{and} \quad
\zeta\frac{d}{d\zeta}F_{f \ot h}(x,\bT;\mu,\zeta)= -\mathcal{D}(\mu,\bT)F_{f\ot h}(x,\bT;\mu,\zeta).
\\ \nn
\mu \frac{d}{d\mu}\mathcal{D}(\mu,\bT)&=&\Gamma(\mu)
\quad \text{and}\quad  \zeta \frac{d}{d\zeta}\gamma_F(\mu,\zeta)=-\Gamma(\mu) ,
\label{th:dDD=G}
\\
\gamma_F(\mu,\zeta)&=&\Gamma(\mu) \ln\(\frac{\mu^2}{\zeta}\)-\gamma_V(\mu),
\end{eqnarray}
and on the right hand side of these equation we have omitted the reference to  flavor $f$ for simplicity. 
The TMD anomalous dimension $\gamma_F(\mu,\zeta)$  contains both single and double logarithms and the anomalous dimension $\gamma_V$ refers to the finite part of the renormalization of the vector form factor, see tab.~\ref{tab:ADs}.  The function $\mathcal{D}(\mu,\bT)$ is the rapidity anomalous dimension, resulting from the TMD factorization of rapidity divergences and actually depends only on $(\mu,|\bT|)$.
It is remarkable that  eq.~(\ref{th:dDD=G}) cannot fix the logarithmic part of $\mathcal{D}$ entirely, but only order by order in  perturbation theory, because the parameter $\mu$ is also responsible for the running of the coupling constant. It has been shown~\cite{Parisi:1979se,Korchemsky:1994is,Scimemi:2016ffw}  that the perturbative series for $\mathcal{D}$ is asymptotical and it has a renormalon pole, whose contribution is significant at large-$b$. Therefore, the rapidity anomalous dimension $\mathcal{D}$ is generically a non-perturbative function, which admits a perturbative expansion only for small values of the parameter $|\bT|$.  One can compare this with the situation in conformal field theory, where the coupling constant is independent on $\mu$, the rapidity anomalous dimension is linear in logarithms of $\mu_b$ and maps to the soft anomalous dimension by conformal transformation~\cite{Vladimirov:2016dll,Vladimirov:2017ksc}.

 The double-evolution equation  of the TMDs can be formulated as in \cite{Scimemi:2018xaf}  using a  two-dimensional vector field notation.
The procedure consists in  introducing a convenient two-dimensional variable which treats scales $\mu$ and $\zeta$ equally,
\begin{eqnarray}\label{def:nu}
\vec \nu=\(\ln (\mu^2/(\text{1 GeV}^2),\ln (\zeta/(1 \text{GeV}^2)\),
\end{eqnarray}
where   the dimension of the scale parameters is  explicitly indicated and   the bold font means the two-dimensional vectors.
 Then one defines the standard vector differential operations in the plane $\vec \nu$, namely, the gradient and the curl
\begin{eqnarray}\label{def:grad+curl}
\vec \nabla=\frac{d}{d\vec \nu}=\(\mu^2\frac{d}{d\mu^2},\zeta\frac{d}{d\zeta}\),\qquad \textbf{curl}=\(-\zeta\frac{d}{d\zeta},\mu^2\frac{d}{d\mu^2}\).
\end{eqnarray}
The TMD anomalous dimensions can be all included in a vector evolution field $\mathbf{E}(\vec \nu,\bT)$,
\begin{eqnarray}
\mathbf{E}(\vec \nu,\bT)=\frac{1}{2}(\gamma_F(\vec \nu),-2\ \mathcal{D}(\vec \nu,\bT)).
\end{eqnarray}
Here and in the following, we use the vectors $\vec \nu$ as the argument of the anomalous dimensions for brevity, keeping in  mind that $\mathcal{D}(\vec \nu,\bT)=\mathcal{D}(\mu,\bT)$, $\gamma_F(\vec \nu)=\gamma_F(\mu,\zeta)$, etc. In other words, the anomalous dimensions are to be evaluated on the corresponding values of $\mu$ and $\zeta$ defined by value of $\vec \nu$ in eq.~(\ref{def:nu}). The TMD evolution equations (\ref{def:TMD_ev_UV}) and the evolution factor $R$ in this notation have the form
\begin{align}\label{def:TMD_ev_nu}
\vec \nabla F(x,\bT;\vec \nu)=\mathbf{E}(\vec \nu,\bT)F(x,\bT;\vec \nu)
\quad \text{and} \quad
\ln R[b,\vec \nu_f\to \vec \nu_i]=\int_P \mathbf{E}\cdot d\vec \nu.
\end{align}
Using this formalism,  eq.~(\ref{th:dDD=G}) are equivalent to the statement that the evolution flow is \textit{irrotational},
\begin{eqnarray}\label{def:irrotational}
\vec \nabla \times \mathbf{E}=0.
\end{eqnarray}
The irrotational vector fields are \textit{conservative} fields, and they can be presented as a gradient of a \textit{scalar potential},
\begin{eqnarray}\label{def:scalarP}
\mathbf{E}(\vec \nu,\bT)=\vec \nabla U(\vec \nu,\bT),
\end{eqnarray}
i.e.  $U$ is the evolution scalar potential for TMD. According to the gradient theorem any line integral of the field $\mathbf{E}$ is path-independent and equals to the difference of values of potential at end-points. Therefore, the solution for the $R$ factor in eq.~(\ref{def:TMD_ev_nu}) is
\begin{align}\label{th:potential-solution}
\ln R[b;\vec \nu_f\to\vec \nu_i]&=U(\vec \nu_f,\bT)-U(\vec \nu_i,\bT)\ ,
\\
\label{th:potential_explicit}
U(\vec \nu,\bT)&=\int^{\nu_{1}} \frac{\Gamma(s)s-\gamma_V(s)}{2}ds-\mathcal{D}(\vec \nu,\bT)\nu_2+\const(\bT),
\end{align}
and $\nu_{1,2}$ are the first and second  components of the vector $\vec \nu$ in eq.~(\ref{def:nu}), and the last term is an arbitrary $b$-dependent function.

 We recall for completeness the perturbative expansions of all these quantities starting from
the running of the coupling constant $a_s=g^2/(4\pi)^2$,
\begin{eqnarray}\label{def:beta}
\mu^2 \frac{d a_s(\mu)}{d\mu^2}=-\beta(a_s),\qquad \beta(a_s)=\sum_{n=0}^\infty a_s^{n+2}(\mu)\beta_n,
\end{eqnarray}
where $\beta_0=\frac{11}{3}C_A-\frac{2}{3}N_f$.
 The ultraviolet anomalous dimensions read
\begin{eqnarray}
\Gamma(\mu)=\sum_{n=0}^\infty a_s^{n+1}(\mu)\Gamma_n,\qquad \gamma_V(\mu)=\sum_{n=1}^\infty a_s^n(\mu)\gamma_n.
\end{eqnarray}
 The leading coefficients in these expansions are $\Gamma_0=4C_F$ and $\gamma_1=-6C_F$ for the quark. In the gluon case, they are $\Gamma_0=4C_A$ and $\gamma_1=-2\beta_0$.  For the collection of higher order terms see e.g. appendix D in~\cite{Echevarria:2016scs}. The perturbative series for the rapidity anomalous dimension $\mathcal{D}$ is
\begin{eqnarray}\label{th:dnk}
\mathcal{D}(\mu,\bT)=\sum_{n=1}^\infty a_s^n(\mu)\sum_{k=0}^n \mathbf{L}_\mu^k d^{(n,k)},
\end{eqnarray}
where  $d^{(n,k)}$ are numbers. Note, that using  eq.~(\ref{th:dDD=G}) the coefficients $d^{(n,k)}$ with $k>0$ are expressed in the terms of $d^{(i,0)}$, $\Gamma_i$ and the coefficients of $\beta$-function. The leading terms of $\mathcal{D}$ are $d^{(1,1)}=\Gamma_0/2$ and $d^{(1,0)}=0$. The explicit expressions for $d^{(n,k)}$ up to $n=3$ can be found in \cite{Vladimirov:2017ksc}.

\subsection{Formal treatment of TMD evolution in the truncated perturbation theory}

The evolution field presented in the previous section  is conservative only  when the full perturbative expansion of the evolution equations is known. 
In practice only a few terms of the evolution are calculated, so that it is important  to understand  in which sense the evolution  field remains conservative.
 Using the Helmholtz decomposition, the  evolution field is split into two parts
\begin{eqnarray}
\mathbf{E}(\vec \nu,\bT)=\tilde{\mathbf{E}}(\vec \nu,\bT)+\mathbf{\Theta}(\vec \nu,\bT).
\end{eqnarray}
The field $\tilde{\mathbf{E}}$ is irrotational, the field $\mathbf{\Theta}$ is divergence-free and they are orthogonal to each other
\begin{eqnarray}
\text{curl}\tilde{\mathbf{E}}=0,\qquad \vec \nabla\cdot \vec \Theta=0,\qquad \tilde{\mathbf{E}}\cdot \mathbf{\Theta}=0,
\end{eqnarray}
with the notation $\text{curl}(\mathbf{curl})=\nabla^2$. 
Then, one  can write 
 the irrotational field $\tilde{\mathbf{E}}$ as the gradient of a scalar potential
\begin{eqnarray}\label{def:Utilde}
\tilde{\mathbf{E}}(\vec \nu,\bT)=\vec \nabla \tilde U(\vec \nu,\bT) ,
\end{eqnarray}
and only this part of the evolution is conservative.

Instead, the divergence-free part in two-dimensions can be written as the vector curl (see eq.~(\ref{def:grad+curl})) of another scalar potential
\begin{eqnarray}
\mathbf{\Theta}(\vec \nu,\bT)=\mathbf{curl}\,V(\vec \nu,\bT).
\end{eqnarray}
The curl of the evolution field can be calculated using the definitions 
(\ref{th:dDD=G}),
\begin{eqnarray}
\text{curl}\mathbf{E}=\text{curl}\mathbf{\Theta}=\frac{\delta \Gamma(\vec \nu,\bT)}{2}\ ,
&\text{with}&\delta \Gamma(\mu,\bT)=\Gamma(\mu)-\mu\frac{d \mathcal{D}(\mu,\bT)}{d\mu}.
\end{eqnarray}
 The function $\delta \Gamma$  can be calculated order by order in perturbation theory.  For instance at order $N$ one finds
\begin{eqnarray}\label{def:deltaG_N}
\delta \Gamma^{(N)}=2\sum_{n=1}^N\sum_{k=0}^n n \bar \beta_{n-1}(a_s)a_s^{n-1} d^{(n,k)}\mathbf{L}_\mu^k,
&\text{where}&\bar \beta_n(a_s)=\beta(a_s)-\sum_{k=0}^{n-1}\beta_k a_s^{k+2} ,
\end{eqnarray}
is the $\beta$-function with first $n$ terms removed.
For instance, we have
\begin{eqnarray}\label{th:dGamma1}
\delta \Gamma^{(1)}&=&\Gamma_0 \beta(a_s)\mathbf{L}_\mu\sim \mathcal{O}(a_s^2\mathbf{L}_\mu),
\\\label{th:dGamma2}
\delta \Gamma^{(2)}&=&\Gamma_0 \bar \beta_1(a_s)\mathbf{L}_\mu+\beta(a_s)a_s\(\Gamma_0 \beta_0 \mathbf{L}_\mu^2+2 \Gamma_1 \mathbf{L}_\mu+4 d^{(2,0)}\)\sim \mathcal{O}(a_s^3\mathbf{L}_\mu^2).
\end{eqnarray}
In these expressions the $\beta$-function is not expanded because in applications it can be of  different perturbative order with respect to the rest of anomalous dimensions.

The immediate consequence of the fact that the evolution field $ \mathbf{E}$ is no more conservative is that  the evolution factor $R[\bT;\vec \nu_f\to\vec \nu_i]$ is dependent on the path chosen to join the  initial and final points $\vec \nu_i,\;\vec \nu_f$ and this  fact introduces a theoretical error  which can be dominant in certain  implementations of the evolution kernels. 
The difference between two  solutions   evaluated  on different paths is
\begin{eqnarray}\label{th:area-relation}
\ln\frac{R[\bT;\{\mu_1,\zeta_1\}\xrightarrow{P_1} \{\mu_2,\zeta_2\}]}{R[\bT;\{\mu_1,\zeta_1\}\xrightarrow{P_2} \{\mu_2,\zeta_2\}]}=
\oint_{P_1 \cup P_2} \mathbf{E}\cdot d\vec \nu=\frac{1}{2}\int_{\Omega(P_1 \cup P_2)}d^2 \nu\, \delta \Gamma(\vec \nu,\bT),
\end{eqnarray}
where ${P_1 \cup P_2}$ is the closed path built from paths $P_1$ and $P_2$ and $\Omega(P_1 \cup P_2)$ is the area surrounded by these paths. 
Using the independence of $\delta \Gamma$ on the variable $\zeta$,  eq.~(\ref{th:area-relation}) becomes
\begin{eqnarray}
\ln \frac{R[\bT;\{\mu_1,\zeta_1\}\xrightarrow{P_1} \{\mu_2,\zeta_2\}]}{R[\bT;\{\mu_1,\zeta_1\}\xrightarrow{P_2} \{\mu_2,\zeta_2\}]}=
\int_{\mu_2}^{\mu_1}\frac{d\mu}{\mu}\delta \Gamma(\mu,\bT)\ln\(\frac{\zeta_1(\mu)}{\zeta_2(\mu)}\),
\end{eqnarray}
where $\zeta_{1,2}(\mu)$ is the $\zeta$-component of the path $P_{1,2}$ at the scale $\mu$. This equation shows that  the difference between paths becomes bigger with largely separated  rapidity scales $\zeta_i$.

\subsection{Restoring path independence of evolution}
The path independence of the evolution is crucial for the implementation of the perturbative formalism, as its absence can derive into uninterpretable extractions of TMDs or  big theoretical errors.  The path independence can be achieved observing that
\begin{align}
\label{eq:intcon}
\mu\frac{d{\cal D}(\mu,\bT)}{d\mu}=-\zeta\frac{d\gamma(\mu,\zeta)}{d\zeta}
\end{align}
should hold order by order in perturbation theory. Once this is realized it is possible to define null-evolution lines in the $(\mu,\zeta)$ plane, which coincide with equipotential lines,
and the evolution of TMD takes place only between  two  different lines.
I resume here two possible  solutions to this problem, following \cite{Scimemi:2018xaf}.

\subsection{Improved $\mathcal{D}$ scenario}
\label{sec:improvedD}
\begin{figure*}[tb]
\centering
\includegraphics[width=\textwidth]{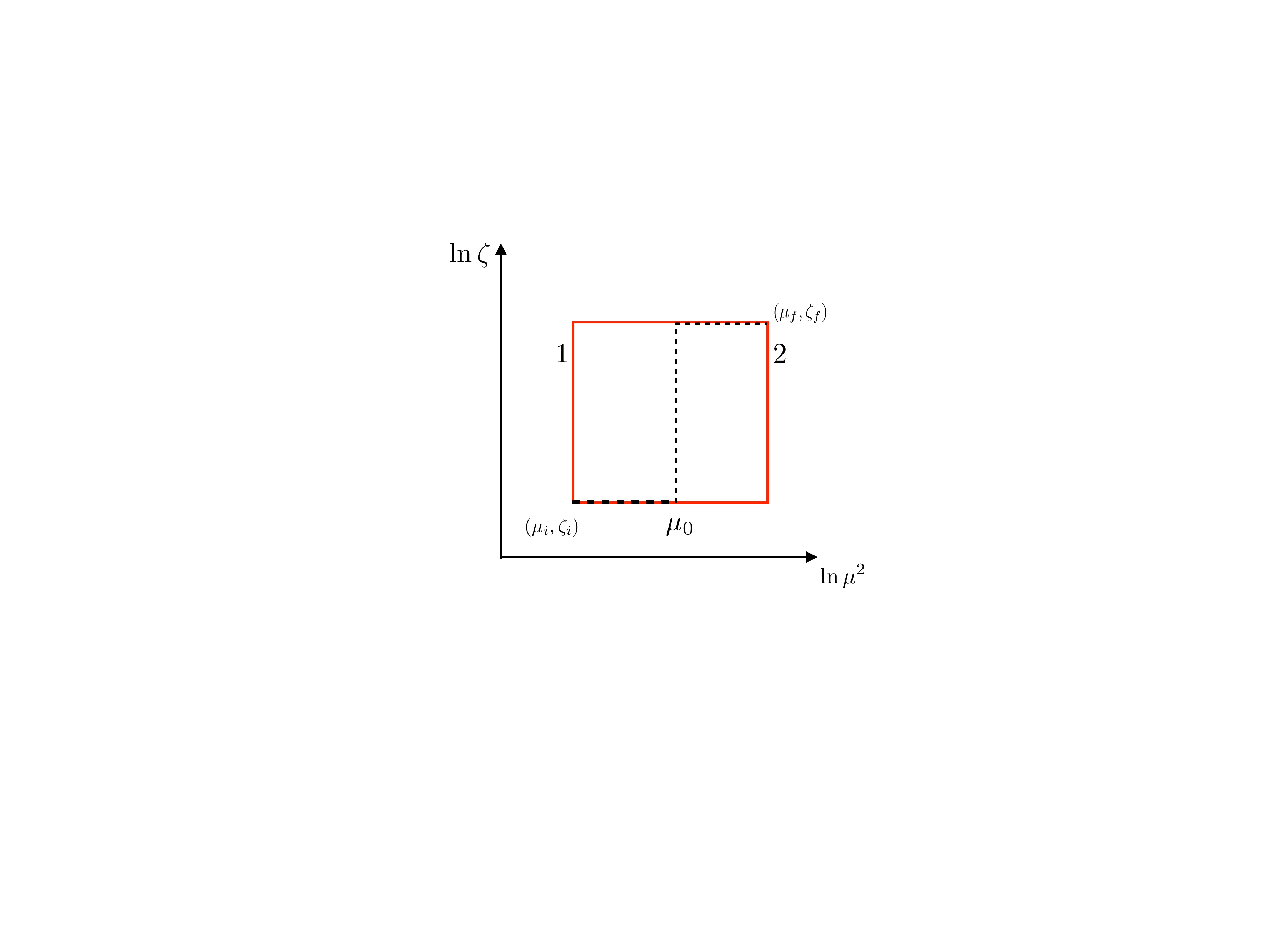}%
\vspace{-4cm}
\caption{Paths  for the improved $\mathcal{D}$ solution  which depend on the choice of the reference scale $\mu_0$.}
\label{fig:paths}
\end{figure*}

In the literature one can find a typical way to implement the evolution that  one can call the improved $\mathcal{D}$ scenario  which  includes the  Collins-Soper-Sterman formalism  \cite{Collins:1981uk,Collins:2011zzd,Aybat:2011zv,Scimemi:2017etj,Echevarria:2012pw,Chiu:2012ir,Li:2016axz}.
In this scenario one chooses a scale $\mu_0$ such that
\begin{eqnarray}\label{imD:mu0}
\delta \Gamma(\mu_0,\bT)=0.
\end{eqnarray} 
 In this  way one obtains
\begin{eqnarray}\label{def:improvedD}
\mathcal{D}(\mu,\bT)=\int^{\mu}_{\mu_0}\frac{d\mu'}{\mu'}\Gamma(\mu')+\mathcal{D}(\mu_0,\bT) ,
\end{eqnarray}
and the scalar potential $\tilde U$ is obtained from eq.~(\ref{th:potential_explicit}) replacing $\mathcal{D}$ by eq.~(\ref{def:improvedD}),
\begin{eqnarray}
\tilde U(\vec \nu,\bT;\mu_0)=\int^{\nu_1}_{\ln \mu^2_0}\frac{\Gamma(s)(s-\nu_2)-\gamma_V(s)}{2}ds-\mathcal{D}(\mu_0,\bT)\nu_2+\const(\bT).
\end{eqnarray}
The TMD evolution factor $R$  depends explicitly on $\mu_0$
\begin{eqnarray}\label{th:RimprovedD}
\text{improved $\mathcal{D}$ solution:}\qquad \ln R[\bT;(\mu_f,\zeta_f)\to(\mu_i,\zeta_i);\mu_0]&=&\int^{\mu_f}_{\mu_i}\frac{d\mu}{\mu}\(\Gamma(\mu)\ln\(\frac{\mu^2}{\zeta_f}\)-\gamma_V(\mu)\)
\\\nn &&-\int_{\mu_0}^{\mu_i}\frac{d\mu}{\mu}\Gamma(\mu)\ln\(\frac{\zeta_f}{\zeta_i}\)-\mathcal{D}(\mu_0,\bT)\ln\(\frac{\zeta_f}{\zeta_i}\).
\end{eqnarray}
The situation in this scenario can be visualized in fig.~\ref{fig:paths}. Choosing a conventional value for $\mu_0$ corresponds to choosing a point where evolution flips from path 1 and path 2 in this  figure. The differences  that can appear in the extraction of TMDs  which depend on the choice of $\mu_0$ can be numerically large, so that the selection of this scale can cause also some problems when a sufficient precision is required.

\subsection{Improved $\gamma$ scenario}
The presence of the intermediate scale $\mu_0$ is not unavoidable in the implementation of the TMD evolution. In fact the integrability condition eq.~(\ref{eq:intcon}) can be restored by changing the anomalous dimension $\gamma_F$ to a modified value $\gamma_M$ such that
\begin{eqnarray}\label{def:improvedGamma}
\gamma_M(\mu,\zeta,\bT)=(\Gamma(\mu)-\delta \Gamma(\mu,\bT))\ln\(\frac{\mu^2}{\zeta}\)-\gamma_V(\mu).
\end{eqnarray}
The corresponding scalar potential $\tilde U$ is derived replacing $\Gamma\to \Gamma-\delta\Gamma$,
\begin{eqnarray}
\tilde U(\vec \nu,\bT)=\int^{\nu_{1}} \frac{(\Gamma(s)-\delta\Gamma(s,\bT))s-\gamma_V(s)}{2}ds-\mathcal{D}(\vec \nu,b_T)\nu_2+\const(\bT).
\end{eqnarray}
Using the definition of $\delta \Gamma$ and integrating by parts one obtains
\begin{eqnarray}
\tilde U(\vec \nu,\bT)=-\int^{\nu_{1}} \(\mathcal{D}(s,\bT)+\frac{\gamma_V(s)}{2}\)ds+\mathcal{D}(\vec \nu,b_T)(\nu_1-\nu_2)+\const(\bT),
\end{eqnarray}
and the corresponding solution for the evolution  factor reads
\begin{eqnarray}\label{th:RimprovedG}
\text{improved $\gamma$ solution:}\qquad \ln R[\bT;(\mu_f,\zeta_f)\to(\mu_i,\zeta_i)]&=&-\int^{\mu_f}_{\mu_i}\frac{d\mu}{\mu}\(2\mathcal{D}(\mu,\bT)+\gamma_V(\mu)\)
\\\nn &&+\mathcal{D}(\mu_f,\bT)\ln\(\frac{\mu_f^2}{\zeta_f}\)-\mathcal{D}(\mu_i,\bT)\ln\(\frac{\mu_i^2}{\zeta_i}\).
\end{eqnarray}
These expressions should be  completed with the resummation of $\mathcal{D}$ by means of renormalization group eq.~(\ref{def:improvedD})  as it is not implicitly included in this scenario.

\subsection{$\zeta$ prescription and optimal TMDs}
\label{subsec:zeta1}

From the discussion of the previous section, and using a correct implementation of TMD evolution it is clear now that TMDs  defined  on  the same equi-potential/null-evolution curves  (that  we call $\vec\omega(\vec \nu_B,b)$) are the same, that is 
\begin{eqnarray}
F(x,\bT;\vec \nu_B)=F(x,\bT;\vec \nu_B'),\qquad \vec \nu_B'\in \vec\omega(\vec \nu_B,\bT),
\end{eqnarray}
when the  scales $\vec \nu_B$ and  $\vec\nu_B'$  belong to the same null-evolution curve. As a consequence  the point  $\vec \nu_B$ in the $(\zeta,\mu)$ plane simply represents a label which defines a null-evolution curve, but it does not enter the function $F(x,b_T;\vec \nu_B)$ explicitly. 
The evolution  of the TMDs occurs only when two TMDs do not belong to the same  null-evolution curve. In this case
\begin{eqnarray}\label{th:Finzeta}
F(x,\bT;\mu_f,\zeta_f)=R[\bT;(\mu_f,\zeta_f)\to (\mu_i,\zeta_{\mu_i}(\vec \nu_B,\bT))]F(x,\bT;\vec \nu_B),
\end{eqnarray}
where $\zeta_\mu$ is defined such that $(\mu_i,\zeta_{\mu_i}(\vec \nu_B,\bT))\in \omega(\vec \nu_B,b_T)$ and $(\mu_f,\zeta_{f})\not \in \omega(\vec \nu_B,\bT)$ .
In order to minimize the evolution effect and so to have a more stable prediction/extraction of TMDs  the initial and final scales should be selected with care.
The final point of the rapidity evolution, $\zeta_f$, is as usual dictated by the hard subprocess.
 On the contrary, the initial value of the rapidity scale $\zeta_i$ should be chosen depending on the input for the  non-perturbative behavior of the TMD distribution. In practice
 it is convenient to  match the TMD distribution to the corresponding collinear distribution. This matching guarantees the agreement of the model to its asymptotic behavior in the limit of high transverse momentum, and determines a significant part of the TMD distribution. The expression for small-$b$ matching has the form
\begin{eqnarray}\label{th:smallB}
F_{f\ot k}(x,\bT;\mu_i,\zeta_i)=\sum_{n}\sum_{f'}C^{(n)}_{f\ot f'}(x,\mathbf{L}_{\mu_i},\mathbf{L}_{\sqrt{\zeta_i}})\otimes f^{(n)}_{f'\ot h}(x,\mu_i),
\end{eqnarray}
where $f$ is PDF or FF, and $C$ is the Wilson coefficient function.
The coefficient function includes the dependence on $\bT$ within the logarithms $\mathbf{L}_\mu$ and $\mathbf{L}_{\sqrt{\zeta}}$. In this way, the initial scales $(\mu_i,\zeta_i)$ explicitly enter in the TMD modeling. Traditionally, see e.g.~\cite{Collins:2011zzd,Aybat:2011zv,Bacchetta:2017gcc},  many studies use $\zeta_i=\mu_i^2$, however  this choice has serious drawbacks, because it leaves uncanceled  logarithmic factors in the coefficient function which get larger and larger at small transverse momentum. More prescriptions are then used to solve this problem, like the $b^*$ prescription \cite{Collins:2011zzd}.  On top of this an eventual usage of  $b^*$ prescription in the evolution factor 
spoils the distinction between the non-perturbative  part of the evolution and the intrinsic  non-perturbative transverse momentum dependence of the partons inside the hadrons.

 The $\zeta$-prescription  suggested in \cite{Scimemi:2017etj,Scimemi:2018xaf} provides an attempt to improve the stability of the perturbative series and 
 {\bf to separate the modeling of the TMD distribution from the factorization procedure}.
  In non-$\zeta$-prescription formulation the TMD distribution has a $\mu$-dependence that is typically related to the scale $b$. Thus the evolution, and hence non-perturbative modification of $\mathcal{D}$, is somehow incorporated into the model for the TMD distribution. This fact makes difficult and sometimes impossible the comparison among different TMD non-perturbative estimations such as lattice or low-energy effective theories.
  Even in  the extraction of TMDs from data one would like to have information on the nonperturbative part of the evolution kernel and the intrinsic nonperturbative TMD initial distribution independently.

The $\zeta$-prescription consists in a special choice of $\zeta_i$ value as a function of $\mu$ and $b$. A TMD distribution in the $\zeta$-prescription reads
\begin{eqnarray}\label{th:Finzeta}
F(x,\bT;\mu_f,\zeta_f)=R[b;(\mu_f,\zeta_f)\to (\mu_i,\zeta_{\mu_i}(\vec \nu_B,\bT))]F(x,b_T;\vec \nu_B),
\end{eqnarray}
where $\zeta_\mu$ is defined such that $(\mu_i,\zeta_{\mu_i}(\vec \nu_B,\bT))\in \omega(\vec \nu_B,\bT)$, that is, the function of $\zeta_\mu(\bT)$ draws a null-evolution curve in the  $(\mu,\zeta)$ plane. 
The value of $\zeta_i$ is selected such that the initial scales TMD distribution $(\mu_i,\zeta_{\mu_i})$ belong to a particular curve.
Note that  in this way we have a line of equivalent initial conditions, provided by all points  $(\mu_i,\zeta_{\mu_i})$  which belong to the same evolution curve.

In order to provide an initial point for the evolution it is  convenient to  re-write eq.~(\ref{th:smallB})  specifying  the scales,
\begin{eqnarray}
F_{f\ot k}(x,\bT;\vec \nu_B)=\sum_{n}\sum_{f'}C^{(n)}_{f\ot f'}(x,\bT,\vec \nu_B,\mu_{\text{OPE}})\otimes f^{(n)}_{f'\ot h}(x,\mu_{\text{OPE}})\ ,
\end{eqnarray}
where $\mu_{\text{OPE}}$ is an intrinsic scale for the expansion of the TMD in terms of Wilson coefficients and PDFs and it is a free parameter.
The values of  $\mu_{\text{OPE}}$  are restricted by the values of $\mu$ spanned by the defining null-evolution curve. In accordance to the general structure of the evolution plane one finds the  following restrictions on the parameter $\mu_{\text{OPE}}$
\begin{eqnarray}
\text{if }\nu_{B,1}<\ln \mu^2_{\text{saddle}}&\Rightarrow & \mu_{\text{OPE}}<\mu_{\text{saddle}},\nn
\\
\text{if }\nu_{B,1}>\ln \mu^2_{\text{saddle}}&\Rightarrow & \mu_{\text{OPE}}>\mu_{\text{saddle}},\nn
\\
\text{if }\vec \nu_{B}=(\ln \mu^2_{\text{saddle}},\ln\zeta_{\text{saddle}})&\Rightarrow & \mu_{\text{OPE}}\text{ unrestricted}.
\end{eqnarray}
It is clear that the last case is preferable, since the model of TMD distribution is completely unrestricted. Additionally, only this case has a unique definition.
The choice of $\mu_{\text{saddle}}$ as the initial point is so optimal and consistent with the  re-expression of TMDs  using PDFs. This choice determines the 
optimal TMD distribution and its related special null-evolution curve. The  definition of the  initial point is therefore non-perturbative, unique and scale-independent.
In such a  way one can denote the optimal TMD simply as $F(x,\bT)$.

\begin{figure}[t]
\centering
\includegraphics[width=0.3\textwidth,angle=270]{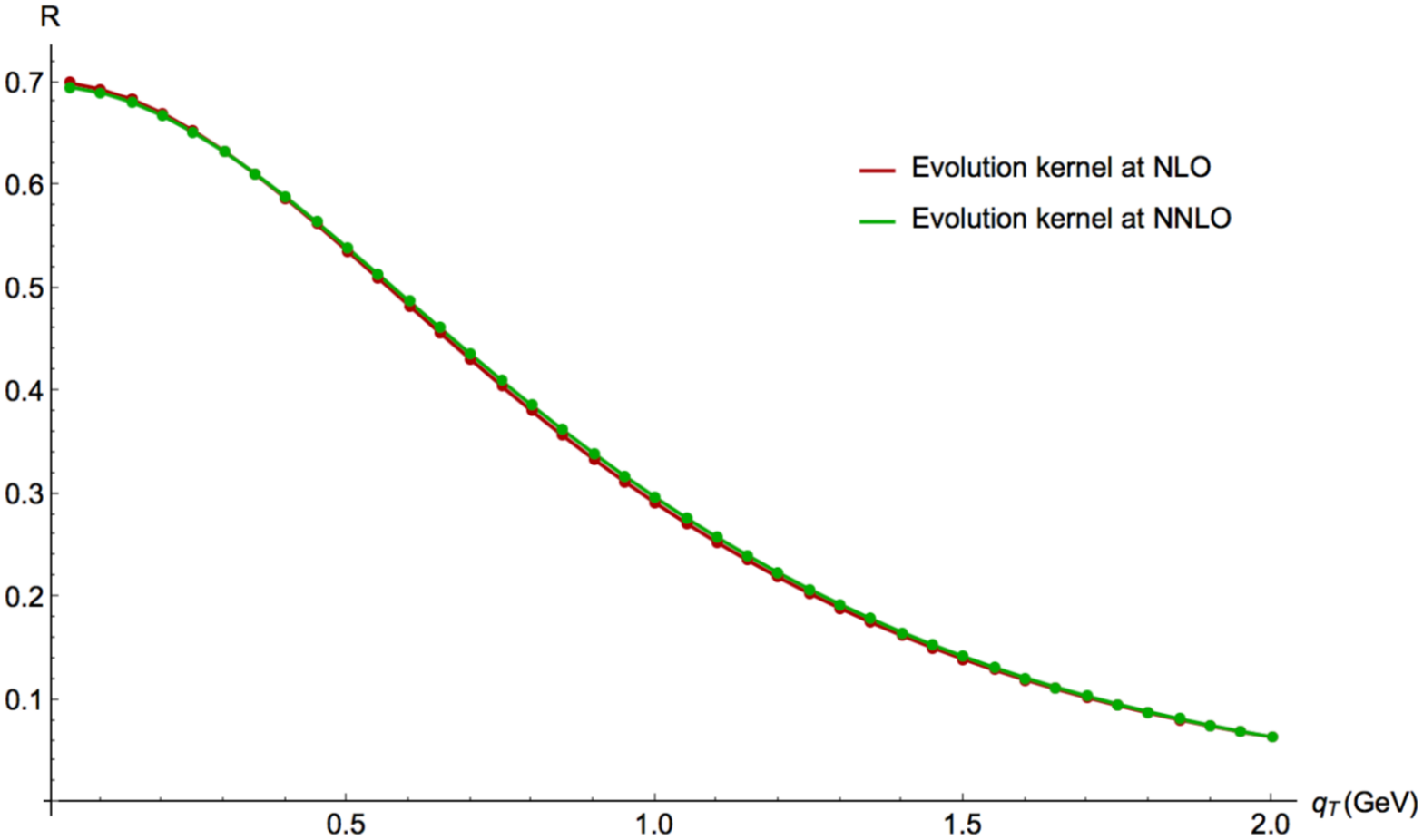}
\includegraphics[width=0.3\textwidth,angle=270]{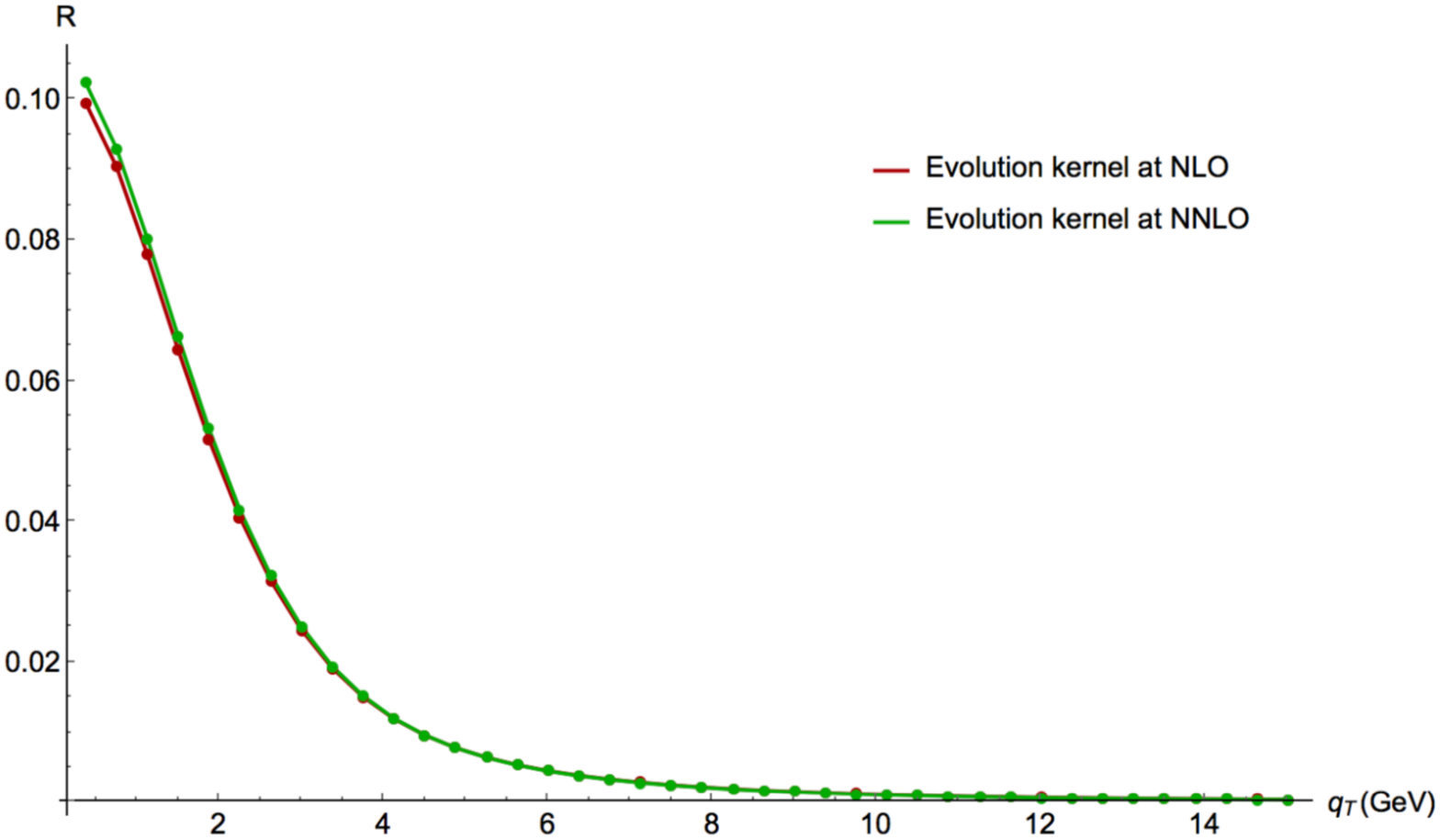}
\caption{$R$ evolution factor using the optimal TMD prescriptions  when  the high scale is fixed at the Z boson mass (right side), and at BELLE center of mass energy 10.52 GeV (left side). In this figure  one has chosen $b_{max}=2.5$ GeV$^{-1}$ and ${\cal D}_{NP}=\exp(-g_K \bT^2)$ with $g_K=0.1$ GeV$^2$.
\label{fig:r}}
\end{figure}

The implementation of the optimal TMD configuration is  compatible with other well-known requests for the evolution and an example of plot for the $R$-factor is given in fig.~\ref{fig:r}.
For instance, at large-$b$ the shape of the rapidity anomalous dimension is non-perturbative  and unknown, which is confirmed also a by an explicit  
renormalon calculation ~\cite{Scimemi:2016ffw,Korchemsky:1994is} (see also~\cite{Becher:2013iya}). 
 So, at large-$b$ the expression for $\mathcal{D}$  should be extracted from data fitting, while at small-$b$ it should match the perturbative expression. 
 In principle  there are several possibilities to account for this effect. For instance one can introduce a simple ansatz like
 the modification
\begin{eqnarray}\label{def:b*}
\mathcal{D}_{\text{NP}}(\mu,b)=\mathcal{D}(\mu,b^*), \qquad b^*(b)=\left\{\begin{array}{cc} b,& b\ll \bar b, \\ b_{\text{max}}, & b\gg \bar b,
\end{array}\right.
\end{eqnarray}
where $b=|\bT|$ and  $b_{\max}$ is a parameter, such that $b_{\text{max}}<\bar b$ as suggested a long ago in~\cite{Collins:1981va},
\begin{eqnarray}
b^*(b)=b\(1+\frac{b^2}{b_{\max}^2}\)^{-1/2},
\end{eqnarray}
as part of the $b^*$ prescription~\cite{Collins:2011zzd}. Let us stress that the choice of a $b^*$ can be admissible separately for the evolution factor and that  eq.~(\ref{def:b*}) does not imply $b^*$-prescription for the whole TMD distribution.
With the choice $b_{\text{max}}<\bar b$ the saddle point is always in the observable region, which  allows to determine the optimal TMD.
The expression for the cross-section with  the optimal TMD definition is particularly compact and reads
\begin{eqnarray}\label{xSec:UNITMD}
\frac{d\sigma}{dX}=\sigma_0 \sum_f \int\frac{d^2 b}{4\pi} e^{i(b\cdot q_T)} H_{ff'}(Q,\mu_f) \{R^f[b;(\mu_f,\zeta_f)]\}^2  F_{f\ot h}(x_1,b)F_{f'\ot h}(x_2,b),
\end{eqnarray}
where the evolution exponent can be given by the equivalent expressions
\begin{eqnarray}\label{xSec:UNIR}
R^f[b;(\mu_f,\zeta_f)]&=&\exp\Big\{-\int^{\mu_f}_{\mu_{\text{saddle}}}\frac{d\mu}{\mu}\(2\mathcal{D}_{\text{NP}}^f(\mu,b)+\gamma^f_V(\mu)\)
+\mathcal{D}_{\text{NP}}^f(\mu_f,b)\ln\(\frac{\mu_f^2}{\zeta_f}\)\Big\}
\\\label{eq:69}&=&\exp\Big\{-\mathcal{D}_{\text{NP}}^f(\mu_f,b)\ln\(\frac{\zeta_f}{\zeta_{\mu_f}(b)}\)\Big\}.
\end{eqnarray}
In eq.~(\ref{xSec:UNIR}), the scale $\mu_{\text{saddle}}$ is $b$-dependent, and defined by the equation
\begin{eqnarray}\label{xSec:musaddle}
\mathcal{D}^f_{\text{NP}}(\mu_{\text{saddle}},b)=0,
\end{eqnarray}
and  the universality of the TMD evolution allows this construction for all types of TMD distribution.
The derivation of the saddle point using formula (\ref{xSec:musaddle})  is in practice done numerically, so that an efficient method to extract it or to approximate  this point should be discussed as in~\cite{Scimemi:2018xaf}.  A technical discussion of this issue is beyond the point of this paper.

\begin{figure}[t]
\centering
\includegraphics[width=0.95\textwidth]{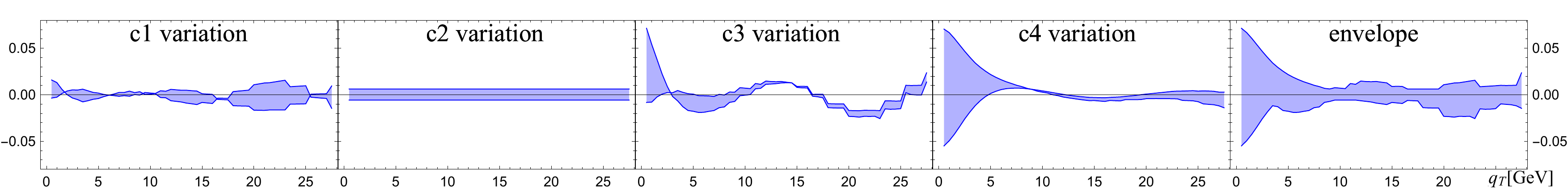}
\includegraphics[width=0.95\textwidth]{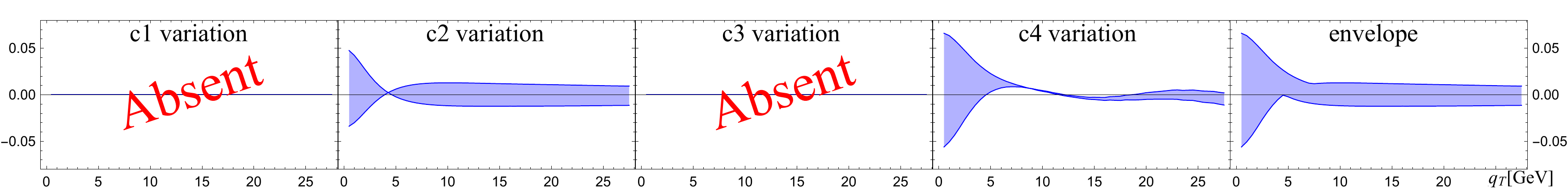}
\caption{Comparison of error bands obtained by the scale-variations for cross-sections at NNLO in traditional (upper figure) and optimal (lower figure) TMD implementations. Here, the kinematics bin-integration, etc., is for the Z-boson production measure at ATLAS at 8 TeV \cite{Aad:2015auj}.} 
\label{fig:ATLAS}
\end{figure}

The optimal TMD definition allows  in general a more self contained and organized discussion of theoretical errors. The absence of an intermediate scale $\mu_0$, remove one artificial source of error while ensuring the path independence of the final result. In this way it is possible to directly compare  $\mathcal{D}_{\text{NP}}$  from different extractions and models. 
 The definition also removes  the question of the low-energy  normalization point $\mu_i$. In  fact  the low-energy normalization is defined "non-perturbatively" and uniquely by eq.~(\ref{xSec:musaddle}).  This fact has the important consequence that the perturbative order of the evolution is completely unrelated to the perturbative order of matching of the TMD on the respective collinear functions. Because the evolution factor is known  often at higher orders then the Wilson coefficient matching factors, it is possible to fully use all the available perturbative information  in  whatsoever  TMD extraction.
 Another important consequence is that in order to compute the theoretical uncertainty of TMDs one is left only with the variation of $\mu_f$ and $\mu_{\rm OPE}$.
 The fact that the number of  varied scales is  different from more standard analysis does not necessarily imply a reduction of theoretical errors. 
 The error in fact reshuffles in 
  $\mu_f$ and $\mu_{\rm OPE}$ but the descriptions appear now more coherent. One can appreciate this effect in  fig.~\ref{fig:ATLAS}.
  In this figure one compares for the ATLAS experiment a 
 standard method to test the dependence on the scales, and thus the stability of the perturbation theory prediction, multiplying each scale by a parameter~\cite{Nadolsky:2000ky,Bozzi:2010xn,DAlesio:2014mrz,Scimemi:2017etj} and varying the parameters nearby their central value. E.g. in the notation of~\cite{Scimemi:2017etj}, one changes scales as
\begin{eqnarray}\label{xSec:c1234}
\mu_0\to c_1 \mu_0,\qquad \mu_f\to c_2 \mu_f,\qquad \mu_i\to c_3 \mu_i,\qquad \mu_{\text{OPE}}\to c_4\mu_{\text{OPE}},
\end{eqnarray}
and checks the variations of $c_i\in(1/2,2)$. The variation of all these four parameters is consistent  with a non-optimal definition of TMDs, while  in the optimal case only the variation of $c_2$ and $c_4$ is necessary.

\section{Conclusions}
The formulation of  factorization theorems in terms of TMDs is a first fundamental step for the study of  the structure of hadrons and  the origin of spin.
The use of the  effective field theory appears  essential  to correctly order the QCD contributions. Properties of TMDs like evolution and their asymptotic limit at large values of transverse momentum can be systematically calculated  starting from the definition of correct operators and the evaluation of the  interesting matrix elements.  A key point  for the renormalization of TMDs is represented  by the so called soft matrix element which is common in the definition of all spin  dependent leading twist  TMD.

Still, all this is just a starting point  for the study of TMDs.
In fact a correct implementation of evolution requires a control of all renormalization scales that  appear in the factorization theorem.  I have  described here some of these possibility  putting the accent  on some recent interesting developments which, at least theoretically allow a better control of the  resummed QCD series.
The understanding of factorization allows also to  precisely define the range of ideal experimental conditions where this formalism can be applied. A  full  analysis of present data using all the theoretical information collected so far is still missing and it will certainly  be an object of research in the forthcoming years.
The formalism described in this work is the one developed for unpolarized distributions. However the evolution factors are  universal, that  is, they are the same for polarized and un-polarized leading twist TMDs and they are valid in  Drell-Yan, SIDIS  experiments and $e^+e^-$ colliders, where the factorization theorem applies.
All this formalism is expected to be tested on data in the near future.  Nevertheless   a lot of perturbative and non-perturbative information is still missing.
Giving a look at tab.~\ref{tab:final-table} one can see that for many TMD one has only a lowest order perturbative calculation which should  be improved  in order to have  a reliable description of data.  
While,
the information  on the non-perturbative structure of TMD is still poor and still driven  by phenomenological models,  it is important to  implement the TMD formalism in such a way that perturbative and non-perturbative effects are well separated. And among the non-perturbative effects,  one should be able to distinguish the ones of the evolution  kernel from the rest.
In the text I have  discussed a possible solution to this problem.
Some prominent research lines which possibly  will deserve  more attention in the future
  include the cases where hadrons are measured inside the jets, see for instance~\cite{Kang:2016ron,Kang:2017glf,Kang:2017btw} or outside a jet (say, hadron-jet interactions)~\cite{Gutierrez-Reyes:2018qez,Liu:2018trl,Buffing:2018ggv} and  lattice.
 
\acknowledgments 
I would like to thank Alexey Vladimirov and Daniel Gutierrez Reyes for discussing  this paper.
I.S. is supported by the Spanish MECD grant FPA2016-75654-C2-2-P.

\bibliography{TMD_ref}

\begin{thebibliography}{95}
\expandafter\ifx\csname natexlab\endcsname\relax\def\natexlab#1{#1}\fi
\expandafter\ifx\csname bibnamefont\endcsname\relax
  \def\bibnamefont#1{#1}\fi
\expandafter\ifx\csname bibfnamefont\endcsname\relax
  \def\bibfnamefont#1{#1}\fi
\expandafter\ifx\csname citenamefont\endcsname\relax
  \def\citenamefont#1{#1}\fi
\expandafter\ifx\csname url\endcsname\relax
  \def\url#1{\texttt{#1}}\fi
\expandafter\ifx\csname urlprefix\endcsname\relax\def\urlprefix{URL }\fi
\providecommand{\bibinfo}[2]{#2}
\providecommand{\eprint}[2][]{\url{#2}}

\bibitem[{\citenamefont{Gribov and Lipatov}(1972)}]{Gribov:1972ri}
\bibinfo{author}{\bibfnamefont{V.~N.} \bibnamefont{Gribov}} \bibnamefont{and}
  \bibinfo{author}{\bibfnamefont{L.~N.} \bibnamefont{Lipatov}},
  \bibinfo{journal}{Sov. J. Nucl. Phys.} \textbf{\bibinfo{volume}{15}},
  \bibinfo{pages}{438} (\bibinfo{year}{1972}), \bibinfo{note}{[Yad.
  Fiz.15,781(1972)]}.

\bibitem[{\citenamefont{Dokshitzer}(1977)}]{Dokshitzer:1977sg}
\bibinfo{author}{\bibfnamefont{Y.~L.} \bibnamefont{Dokshitzer}},
  \bibinfo{journal}{Sov. Phys. JETP} \textbf{\bibinfo{volume}{46}},
  \bibinfo{pages}{641} (\bibinfo{year}{1977}), \bibinfo{note}{[Zh. Eksp. Teor.
  Fiz.73,1216(1977)]}.

\bibitem[{\citenamefont{Altarelli and Parisi}(1977)}]{Altarelli:1977zs}
\bibinfo{author}{\bibfnamefont{G.}~\bibnamefont{Altarelli}} \bibnamefont{and}
  \bibinfo{author}{\bibfnamefont{G.}~\bibnamefont{Parisi}},
  \bibinfo{journal}{Nucl. Phys.} \textbf{\bibinfo{volume}{B126}},
  \bibinfo{pages}{298} (\bibinfo{year}{1977}).

\bibitem[{\citenamefont{Parisi and Petronzio}(1979)}]{Parisi:1979se}
\bibinfo{author}{\bibfnamefont{G.}~\bibnamefont{Parisi}} \bibnamefont{and}
  \bibinfo{author}{\bibfnamefont{R.}~\bibnamefont{Petronzio}},
  \bibinfo{journal}{Nucl. Phys.} \textbf{\bibinfo{volume}{B154}},
  \bibinfo{pages}{427} (\bibinfo{year}{1979}).

\bibitem[{\citenamefont{Collins and Soper}(1982)}]{Collins:1981va}
\bibinfo{author}{\bibfnamefont{J.~C.} \bibnamefont{Collins}} \bibnamefont{and}
  \bibinfo{author}{\bibfnamefont{D.~E.} \bibnamefont{Soper}},
  \bibinfo{journal}{Nucl. Phys.} \textbf{\bibinfo{volume}{B197}},
  \bibinfo{pages}{446} (\bibinfo{year}{1982}).

\bibitem[{\citenamefont{Collins et~al.}(1985)\citenamefont{Collins, Soper, and
  Sterman}}]{Collins:1984kg}
\bibinfo{author}{\bibfnamefont{J.~C.} \bibnamefont{Collins}},
  \bibinfo{author}{\bibfnamefont{D.~E.} \bibnamefont{Soper}}, \bibnamefont{and}
  \bibinfo{author}{\bibfnamefont{G.~F.} \bibnamefont{Sterman}},
  \bibinfo{journal}{Nucl. Phys.} \textbf{\bibinfo{volume}{B250}},
  \bibinfo{pages}{199} (\bibinfo{year}{1985}).

\bibitem[{\citenamefont{Ji et~al.}(2005)\citenamefont{Ji, Ma, and
  Yuan}}]{Ji:2004wu}
\bibinfo{author}{\bibfnamefont{X.-d.} \bibnamefont{Ji}},
  \bibinfo{author}{\bibfnamefont{J.-p.} \bibnamefont{Ma}}, \bibnamefont{and}
  \bibinfo{author}{\bibfnamefont{F.}~\bibnamefont{Yuan}},
  \bibinfo{journal}{Phys. Rev.} \textbf{\bibinfo{volume}{D71}},
  \bibinfo{pages}{034005} (\bibinfo{year}{2005}), \eprint{hep-ph/0404183}.

\bibitem[{\citenamefont{Becher and Neubert}(2011)}]{Becher:2010tm}
\bibinfo{author}{\bibfnamefont{T.}~\bibnamefont{Becher}} \bibnamefont{and}
  \bibinfo{author}{\bibfnamefont{M.}~\bibnamefont{Neubert}},
  \bibinfo{journal}{Eur. Phys. J.} \textbf{\bibinfo{volume}{C71}},
  \bibinfo{pages}{1665} (\bibinfo{year}{2011}), \eprint{1007.4005}.

\bibitem[{\citenamefont{Collins}(2013)}]{Collins:2011zzd}
\bibinfo{author}{\bibfnamefont{J.}~\bibnamefont{Collins}},
  \emph{\bibinfo{title}{{Foundations of perturbative QCD}}}
  (\bibinfo{publisher}{Cambridge University Press}, \bibinfo{year}{2013}), ISBN
  \bibinfo{isbn}{9781107645257, 9781107645257, 9780521855334, 9781139097826},
  \urlprefix\url{http://www.cambridge.org/de/knowledge/isbn/item5756723}.

\bibitem[{\citenamefont{Echevarria et~al.}(2012)\citenamefont{Echevarria,
  Idilbi, and Scimemi}}]{GarciaEchevarria:2011rb}
\bibinfo{author}{\bibfnamefont{M.~G.} \bibnamefont{Echevarria}},
  \bibinfo{author}{\bibfnamefont{A.}~\bibnamefont{Idilbi}}, \bibnamefont{and}
  \bibinfo{author}{\bibfnamefont{I.}~\bibnamefont{Scimemi}},
  \bibinfo{journal}{JHEP} \textbf{\bibinfo{volume}{07}}, \bibinfo{pages}{002}
  (\bibinfo{year}{2012}), \eprint{1111.4996}.

\bibitem[{\citenamefont{Chiu et~al.}(2012{\natexlab{a}})\citenamefont{Chiu,
  Jain, Neill, and Rothstein}}]{Chiu:2012ir}
\bibinfo{author}{\bibfnamefont{J.-Y.} \bibnamefont{Chiu}},
  \bibinfo{author}{\bibfnamefont{A.}~\bibnamefont{Jain}},
  \bibinfo{author}{\bibfnamefont{D.}~\bibnamefont{Neill}}, \bibnamefont{and}
  \bibinfo{author}{\bibfnamefont{I.~Z.} \bibnamefont{Rothstein}},
  \bibinfo{journal}{JHEP} \textbf{\bibinfo{volume}{05}}, \bibinfo{pages}{084}
  (\bibinfo{year}{2012}{\natexlab{a}}), \eprint{1202.0814}.

\bibitem[{\citenamefont{Echevarria
  et~al.}(2013{\natexlab{a}})\citenamefont{Echevarria, Idilbi, and
  Scimemi}}]{Echevarria:2012js}
\bibinfo{author}{\bibfnamefont{M.~G.} \bibnamefont{Echevarria}},
  \bibinfo{author}{\bibfnamefont{A.}~\bibnamefont{Idilbi}}, \bibnamefont{and}
  \bibinfo{author}{\bibfnamefont{I.}~\bibnamefont{Scimemi}},
  \bibinfo{journal}{Phys. Lett.} \textbf{\bibinfo{volume}{B726}},
  \bibinfo{pages}{795} (\bibinfo{year}{2013}{\natexlab{a}}),
  \eprint{1211.1947}.

\bibitem[{\citenamefont{Aybat and Rogers}(2011)}]{Aybat:2011zv}
\bibinfo{author}{\bibfnamefont{S.~M.} \bibnamefont{Aybat}} \bibnamefont{and}
  \bibinfo{author}{\bibfnamefont{T.~C.} \bibnamefont{Rogers}},
  \bibinfo{journal}{Phys. Rev.} \textbf{\bibinfo{volume}{D83}},
  \bibinfo{pages}{114042} (\bibinfo{year}{2011}), \eprint{1101.5057}.

\bibitem[{\citenamefont{Echevarria
  et~al.}(2013{\natexlab{b}})\citenamefont{Echevarria, Idilbi, Schafer, and
  Scimemi}}]{Echevarria:2012pw}
\bibinfo{author}{\bibfnamefont{M.~G.} \bibnamefont{Echevarria}},
  \bibinfo{author}{\bibfnamefont{A.}~\bibnamefont{Idilbi}},
  \bibinfo{author}{\bibfnamefont{A.}~\bibnamefont{Schafer}}, \bibnamefont{and}
  \bibinfo{author}{\bibfnamefont{I.}~\bibnamefont{Scimemi}},
  \bibinfo{journal}{Eur. Phys. J.} \textbf{\bibinfo{volume}{C73}},
  \bibinfo{pages}{2636} (\bibinfo{year}{2013}{\natexlab{b}}),
  \eprint{1208.1281}.

\bibitem[{\citenamefont{D'Alesio et~al.}(2014)\citenamefont{D'Alesio,
  Echevarria, Melis, and Scimemi}}]{DAlesio:2014mrz}
\bibinfo{author}{\bibfnamefont{U.}~\bibnamefont{D'Alesio}},
  \bibinfo{author}{\bibfnamefont{M.~G.} \bibnamefont{Echevarria}},
  \bibinfo{author}{\bibfnamefont{S.}~\bibnamefont{Melis}}, \bibnamefont{and}
  \bibinfo{author}{\bibfnamefont{I.}~\bibnamefont{Scimemi}},
  \bibinfo{journal}{JHEP} \textbf{\bibinfo{volume}{11}}, \bibinfo{pages}{098}
  (\bibinfo{year}{2014}), \eprint{1407.3311}.

\bibitem[{\citenamefont{Scimemi and
  Vladimirov}(2018{\natexlab{a}})}]{Scimemi:2017etj}
\bibinfo{author}{\bibfnamefont{I.}~\bibnamefont{Scimemi}} \bibnamefont{and}
  \bibinfo{author}{\bibfnamefont{A.}~\bibnamefont{Vladimirov}},
  \bibinfo{journal}{Eur. Phys. J.} \textbf{\bibinfo{volume}{C78}},
  \bibinfo{pages}{89} (\bibinfo{year}{2018}{\natexlab{a}}),
  \eprint{1706.01473}.

\bibitem[{\citenamefont{Scimemi and
  Vladimirov}(2018{\natexlab{b}})}]{Scimemi:2018xaf}
\bibinfo{author}{\bibfnamefont{I.}~\bibnamefont{Scimemi}} \bibnamefont{and}
  \bibinfo{author}{\bibfnamefont{A.}~\bibnamefont{Vladimirov}},
  \bibinfo{journal}{JHEP} \textbf{\bibinfo{volume}{08}}, \bibinfo{pages}{003}
  (\bibinfo{year}{2018}{\natexlab{b}}), \eprint{1803.11089}.

\bibitem[{\citenamefont{Idilbi and
  Scimemi}(2011{\natexlab{a}})}]{Idilbi:2010im}
\bibinfo{author}{\bibfnamefont{A.}~\bibnamefont{Idilbi}} \bibnamefont{and}
  \bibinfo{author}{\bibfnamefont{I.}~\bibnamefont{Scimemi}},
  \bibinfo{journal}{Phys. Lett.} \textbf{\bibinfo{volume}{B695}},
  \bibinfo{pages}{463} (\bibinfo{year}{2011}{\natexlab{a}}),
  \eprint{1009.2776}.

\bibitem[{\citenamefont{Idilbi and
  Scimemi}(2011{\natexlab{b}})}]{Idilbi:2010tc}
\bibinfo{author}{\bibfnamefont{A.}~\bibnamefont{Idilbi}} \bibnamefont{and}
  \bibinfo{author}{\bibfnamefont{I.}~\bibnamefont{Scimemi}},
  \bibinfo{journal}{AIP Conf. Proc.} \textbf{\bibinfo{volume}{1343}},
  \bibinfo{pages}{320} (\bibinfo{year}{2011}{\natexlab{b}}),
  \eprint{1012.4419}.

\bibitem[{\citenamefont{Garcia-Echevarria
  et~al.}(2011)\citenamefont{Garcia-Echevarria, Idilbi, and
  Scimemi}}]{GarciaEchevarria:2011md}
\bibinfo{author}{\bibfnamefont{M.}~\bibnamefont{Garcia-Echevarria}},
  \bibinfo{author}{\bibfnamefont{A.}~\bibnamefont{Idilbi}}, \bibnamefont{and}
  \bibinfo{author}{\bibfnamefont{I.}~\bibnamefont{Scimemi}},
  \bibinfo{journal}{Phys. Rev.} \textbf{\bibinfo{volume}{D84}},
  \bibinfo{pages}{011502} (\bibinfo{year}{2011}), \eprint{1104.0686}.

\bibitem[{\citenamefont{Vladimirov}(2016)}]{Vladimirov:2016qkd}
\bibinfo{author}{\bibfnamefont{A.}~\bibnamefont{Vladimirov}},
  \bibinfo{journal}{JHEP} \textbf{\bibinfo{volume}{12}}, \bibinfo{pages}{038}
  (\bibinfo{year}{2016}), \eprint{1608.04920}.

\bibitem[{\citenamefont{Vladimirov}(2018)}]{Vladimirov:2017ksc}
\bibinfo{author}{\bibfnamefont{A.}~\bibnamefont{Vladimirov}},
  \bibinfo{journal}{JHEP} \textbf{\bibinfo{volume}{04}}, \bibinfo{pages}{045}
  (\bibinfo{year}{2018}), \eprint{1707.07606}.

\bibitem[{\citenamefont{Echevarria
  et~al.}(2014{\natexlab{a}})\citenamefont{Echevarria, Idilbi, and
  Scimemi}}]{Echevarria:2013aca}
\bibinfo{author}{\bibfnamefont{M.~G.} \bibnamefont{Echevarria}},
  \bibinfo{author}{\bibfnamefont{A.}~\bibnamefont{Idilbi}}, \bibnamefont{and}
  \bibinfo{author}{\bibfnamefont{I.}~\bibnamefont{Scimemi}},
  \bibinfo{journal}{Int. J. Mod. Phys. Conf. Ser.}
  \textbf{\bibinfo{volume}{25}}, \bibinfo{pages}{1460005}
  (\bibinfo{year}{2014}{\natexlab{a}}), \eprint{1310.8541}.

\bibitem[{\citenamefont{Echevarria
  et~al.}(2016{\natexlab{a}})\citenamefont{Echevarria, Scimemi, and
  Vladimirov}}]{Echevarria:2015byo}
\bibinfo{author}{\bibfnamefont{M.~G.} \bibnamefont{Echevarria}},
  \bibinfo{author}{\bibfnamefont{I.}~\bibnamefont{Scimemi}}, \bibnamefont{and}
  \bibinfo{author}{\bibfnamefont{A.}~\bibnamefont{Vladimirov}},
  \bibinfo{journal}{Phys. Rev.} \textbf{\bibinfo{volume}{D93}},
  \bibinfo{pages}{054004} (\bibinfo{year}{2016}{\natexlab{a}}),
  \eprint{1511.05590}.

\bibitem[{\citenamefont{Echevarria
  et~al.}(2016{\natexlab{b}})\citenamefont{Echevarria, Scimemi, and
  Vladimirov}}]{Echevarria:2016scs}
\bibinfo{author}{\bibfnamefont{M.~G.} \bibnamefont{Echevarria}},
  \bibinfo{author}{\bibfnamefont{I.}~\bibnamefont{Scimemi}}, \bibnamefont{and}
  \bibinfo{author}{\bibfnamefont{A.}~\bibnamefont{Vladimirov}},
  \bibinfo{journal}{JHEP} \textbf{\bibinfo{volume}{09}}, \bibinfo{pages}{004}
  (\bibinfo{year}{2016}{\natexlab{b}}), \eprint{1604.07869}.

\bibitem[{\citenamefont{Gutierrez-Reyes
  et~al.}(2017)\citenamefont{Gutierrez-Reyes, Scimemi, and
  Vladimirov}}]{Gutierrez-Reyes:2017glx}
\bibinfo{author}{\bibfnamefont{D.}~\bibnamefont{Gutierrez-Reyes}},
  \bibinfo{author}{\bibfnamefont{I.}~\bibnamefont{Scimemi}}, \bibnamefont{and}
  \bibinfo{author}{\bibfnamefont{A.~A.} \bibnamefont{Vladimirov}},
  \bibinfo{journal}{Phys. Lett.} \textbf{\bibinfo{volume}{B769}},
  \bibinfo{pages}{84} (\bibinfo{year}{2017}), \eprint{1702.06558}.

\bibitem[{\citenamefont{Gutierrez-Reyes
  et~al.}(2018{\natexlab{a}})\citenamefont{Gutierrez-Reyes, Scimemi, and
  Vladimirov}}]{Gutierrez-Reyes:2018iod}
\bibinfo{author}{\bibfnamefont{D.}~\bibnamefont{Gutierrez-Reyes}},
  \bibinfo{author}{\bibfnamefont{I.}~\bibnamefont{Scimemi}}, \bibnamefont{and}
  \bibinfo{author}{\bibfnamefont{A.}~\bibnamefont{Vladimirov}},
  \bibinfo{journal}{JHEP} \textbf{\bibinfo{volume}{07}}, \bibinfo{pages}{172}
  (\bibinfo{year}{2018}{\natexlab{a}}), \eprint{1805.07243}.

\bibitem[{\citenamefont{Hagler et~al.}(2009)\citenamefont{Hagler, Musch,
  Negele, and Schafer}}]{Hagler:2009mb}
\bibinfo{author}{\bibfnamefont{P.}~\bibnamefont{Hagler}},
  \bibinfo{author}{\bibfnamefont{B.~U.} \bibnamefont{Musch}},
  \bibinfo{author}{\bibfnamefont{J.~W.} \bibnamefont{Negele}},
  \bibnamefont{and} \bibinfo{author}{\bibfnamefont{A.}~\bibnamefont{Schafer}},
  \bibinfo{journal}{EPL} \textbf{\bibinfo{volume}{88}}, \bibinfo{pages}{61001}
  (\bibinfo{year}{2009}), \eprint{0908.1283}.

\bibitem[{\citenamefont{Musch et~al.}(2011)\citenamefont{Musch, Hagler, Negele,
  and Schafer}}]{Musch:2010ka}
\bibinfo{author}{\bibfnamefont{B.~U.} \bibnamefont{Musch}},
  \bibinfo{author}{\bibfnamefont{P.}~\bibnamefont{Hagler}},
  \bibinfo{author}{\bibfnamefont{J.~W.} \bibnamefont{Negele}},
  \bibnamefont{and} \bibinfo{author}{\bibfnamefont{A.}~\bibnamefont{Schafer}},
  \bibinfo{journal}{Phys. Rev.} \textbf{\bibinfo{volume}{D83}},
  \bibinfo{pages}{094507} (\bibinfo{year}{2011}), \eprint{1011.1213}.

\bibitem[{\citenamefont{Musch et~al.}(2012)\citenamefont{Musch, Hagler,
  Engelhardt, Negele, and Schafer}}]{Musch:2011er}
\bibinfo{author}{\bibfnamefont{B.~U.} \bibnamefont{Musch}},
  \bibinfo{author}{\bibfnamefont{P.}~\bibnamefont{Hagler}},
  \bibinfo{author}{\bibfnamefont{M.}~\bibnamefont{Engelhardt}},
  \bibinfo{author}{\bibfnamefont{J.~W.} \bibnamefont{Negele}},
  \bibnamefont{and} \bibinfo{author}{\bibfnamefont{A.}~\bibnamefont{Schafer}},
  \bibinfo{journal}{Phys. Rev.} \textbf{\bibinfo{volume}{D85}},
  \bibinfo{pages}{094510} (\bibinfo{year}{2012}), \eprint{1111.4249}.

\bibitem[{\citenamefont{Ji}(2013)}]{Ji:2013dva}
\bibinfo{author}{\bibfnamefont{X.}~\bibnamefont{Ji}}, \bibinfo{journal}{Phys.
  Rev. Lett.} \textbf{\bibinfo{volume}{110}}, \bibinfo{pages}{262002}
  (\bibinfo{year}{2013}), \eprint{1305.1539}.

\bibitem[{\citenamefont{Ji}(2014)}]{Ji:2014gla}
\bibinfo{author}{\bibfnamefont{X.}~\bibnamefont{Ji}}, \bibinfo{journal}{Sci.
  China Phys. Mech. Astron.} \textbf{\bibinfo{volume}{57}},
  \bibinfo{pages}{1407} (\bibinfo{year}{2014}), \eprint{1404.6680}.

\bibitem[{\citenamefont{Engelhardt et~al.}(2016)\citenamefont{Engelhardt,
  Haegler, Musch, Negele, and Schäfer}}]{Engelhardt:2015xja}
\bibinfo{author}{\bibfnamefont{M.}~\bibnamefont{Engelhardt}},
  \bibinfo{author}{\bibfnamefont{P.}~\bibnamefont{Haegler}},
  \bibinfo{author}{\bibfnamefont{B.}~\bibnamefont{Musch}},
  \bibinfo{author}{\bibfnamefont{J.}~\bibnamefont{Negele}}, \bibnamefont{and}
  \bibinfo{author}{\bibfnamefont{A.}~\bibnamefont{Schäfer}},
  \bibinfo{journal}{Phys. Rev.} \textbf{\bibinfo{volume}{D93}},
  \bibinfo{pages}{054501} (\bibinfo{year}{2016}), \eprint{1506.07826}.

\bibitem[{\citenamefont{Yoon et~al.}(2015)\citenamefont{Yoon, Bhattacharya,
  Engelhardt, Green, Gupta, Hägler, Musch, Negele, Pochinsky, and
  Syritsyn}}]{Yoon:2016dyh}
\bibinfo{author}{\bibfnamefont{B.}~\bibnamefont{Yoon}},
  \bibinfo{author}{\bibfnamefont{T.}~\bibnamefont{Bhattacharya}},
  \bibinfo{author}{\bibfnamefont{M.}~\bibnamefont{Engelhardt}},
  \bibinfo{author}{\bibfnamefont{J.}~\bibnamefont{Green}},
  \bibinfo{author}{\bibfnamefont{R.}~\bibnamefont{Gupta}},
  \bibinfo{author}{\bibfnamefont{P.}~\bibnamefont{Hägler}},
  \bibinfo{author}{\bibfnamefont{B.}~\bibnamefont{Musch}},
  \bibinfo{author}{\bibfnamefont{J.}~\bibnamefont{Negele}},
  \bibinfo{author}{\bibfnamefont{A.}~\bibnamefont{Pochinsky}},
  \bibnamefont{and} \bibinfo{author}{\bibfnamefont{S.}~\bibnamefont{Syritsyn}},
  in \emph{\bibinfo{booktitle}{{Proceedings, 33rd International Symposium on
  Lattice Field Theory (Lattice 2015): Kobe, Japan, July 14-18, 2015}}},
  \bibinfo{organization}{SISSA} (\bibinfo{publisher}{SISSA},
  \bibinfo{year}{2015}), \eprint{1601.05717}.

\bibitem[{\citenamefont{Yoon et~al.}(2017)\citenamefont{Yoon, Engelhardt,
  Gupta, Bhattacharya, Green, Musch, Negele, Pochinsky, Schäfer, and
  Syritsyn}}]{Yoon:2017qzo}
\bibinfo{author}{\bibfnamefont{B.}~\bibnamefont{Yoon}},
  \bibinfo{author}{\bibfnamefont{M.}~\bibnamefont{Engelhardt}},
  \bibinfo{author}{\bibfnamefont{R.}~\bibnamefont{Gupta}},
  \bibinfo{author}{\bibfnamefont{T.}~\bibnamefont{Bhattacharya}},
  \bibinfo{author}{\bibfnamefont{J.~R.} \bibnamefont{Green}},
  \bibinfo{author}{\bibfnamefont{B.~U.} \bibnamefont{Musch}},
  \bibinfo{author}{\bibfnamefont{J.~W.} \bibnamefont{Negele}},
  \bibinfo{author}{\bibfnamefont{A.~V.} \bibnamefont{Pochinsky}},
  \bibinfo{author}{\bibfnamefont{A.}~\bibnamefont{Schäfer}}, \bibnamefont{and}
  \bibinfo{author}{\bibfnamefont{S.~N.} \bibnamefont{Syritsyn}},
  \bibinfo{journal}{Phys. Rev.} \textbf{\bibinfo{volume}{D96}},
  \bibinfo{pages}{094508} (\bibinfo{year}{2017}), \eprint{1706.03406}.

\bibitem[{\citenamefont{Radyushkin}(2017)}]{Radyushkin:2017cyf}
\bibinfo{author}{\bibfnamefont{A.~V.} \bibnamefont{Radyushkin}},
  \bibinfo{journal}{Phys. Rev.} \textbf{\bibinfo{volume}{D96}},
  \bibinfo{pages}{034025} (\bibinfo{year}{2017}), \eprint{1705.01488}.

\bibitem[{\citenamefont{Orginos et~al.}(2017)\citenamefont{Orginos, Radyushkin,
  Karpie, and Zafeiropoulos}}]{Orginos:2017kos}
\bibinfo{author}{\bibfnamefont{K.}~\bibnamefont{Orginos}},
  \bibinfo{author}{\bibfnamefont{A.}~\bibnamefont{Radyushkin}},
  \bibinfo{author}{\bibfnamefont{J.}~\bibnamefont{Karpie}}, \bibnamefont{and}
  \bibinfo{author}{\bibfnamefont{S.}~\bibnamefont{Zafeiropoulos}},
  \bibinfo{journal}{Phys. Rev.} \textbf{\bibinfo{volume}{D96}},
  \bibinfo{pages}{094503} (\bibinfo{year}{2017}), \eprint{1706.05373}.

\bibitem[{\citenamefont{Ji et~al.}(2018)\citenamefont{Ji, Jin, Yuan, Zhang, and
  Zhao}}]{Ji:2018hvs}
\bibinfo{author}{\bibfnamefont{X.}~\bibnamefont{Ji}},
  \bibinfo{author}{\bibfnamefont{L.-C.} \bibnamefont{Jin}},
  \bibinfo{author}{\bibfnamefont{F.}~\bibnamefont{Yuan}},
  \bibinfo{author}{\bibfnamefont{J.-H.} \bibnamefont{Zhang}}, \bibnamefont{and}
  \bibinfo{author}{\bibfnamefont{Y.}~\bibnamefont{Zhao}}
  (\bibinfo{year}{2018}), \eprint{1801.05930}.

\bibitem[{\citenamefont{Manohar and Stewart}(2007)}]{Manohar:2006nz}
\bibinfo{author}{\bibfnamefont{A.~V.} \bibnamefont{Manohar}} \bibnamefont{and}
  \bibinfo{author}{\bibfnamefont{I.~W.} \bibnamefont{Stewart}},
  \bibinfo{journal}{Phys. Rev.} \textbf{\bibinfo{volume}{D76}},
  \bibinfo{pages}{074002} (\bibinfo{year}{2007}), \eprint{hep-ph/0605001}.

\bibitem[{\citenamefont{Ebert et~al.}(2018{\natexlab{a}})\citenamefont{Ebert,
  Stewart, and Zhao}}]{Ebert:2018gzl}
\bibinfo{author}{\bibfnamefont{M.~A.} \bibnamefont{Ebert}},
  \bibinfo{author}{\bibfnamefont{I.~W.} \bibnamefont{Stewart}},
  \bibnamefont{and} \bibinfo{author}{\bibfnamefont{Y.}~\bibnamefont{Zhao}}
  (\bibinfo{year}{2018}{\natexlab{a}}), \eprint{1811.00026}.

\bibitem[{\citenamefont{Scimemi and Vladimirov}(2017)}]{Scimemi:2016ffw}
\bibinfo{author}{\bibfnamefont{I.}~\bibnamefont{Scimemi}} \bibnamefont{and}
  \bibinfo{author}{\bibfnamefont{A.}~\bibnamefont{Vladimirov}},
  \bibinfo{journal}{JHEP} \textbf{\bibinfo{volume}{03}}, \bibinfo{pages}{002}
  (\bibinfo{year}{2017}), \eprint{1609.06047}.

\bibitem[{\citenamefont{Luebbert et~al.}(2016)\citenamefont{Luebbert, Oredsson,
  and Stahlhofen}}]{Luebbert:2016itl}
\bibinfo{author}{\bibfnamefont{T.}~\bibnamefont{Luebbert}},
  \bibinfo{author}{\bibfnamefont{J.}~\bibnamefont{Oredsson}}, \bibnamefont{and}
  \bibinfo{author}{\bibfnamefont{M.}~\bibnamefont{Stahlhofen}},
  \bibinfo{journal}{JHEP} \textbf{\bibinfo{volume}{03}}, \bibinfo{pages}{168}
  (\bibinfo{year}{2016}), \eprint{1602.01829}.

\bibitem[{\citenamefont{Chiu et~al.}(2012{\natexlab{b}})\citenamefont{Chiu,
  Jain, Neill, and Rothstein}}]{Chiu:2011qc}
\bibinfo{author}{\bibfnamefont{J.-y.} \bibnamefont{Chiu}},
  \bibinfo{author}{\bibfnamefont{A.}~\bibnamefont{Jain}},
  \bibinfo{author}{\bibfnamefont{D.}~\bibnamefont{Neill}}, \bibnamefont{and}
  \bibinfo{author}{\bibfnamefont{I.~Z.} \bibnamefont{Rothstein}},
  \bibinfo{journal}{Phys. Rev. Lett.} \textbf{\bibinfo{volume}{108}},
  \bibinfo{pages}{151601} (\bibinfo{year}{2012}{\natexlab{b}}),
  \eprint{1104.0881}.

\bibitem[{\citenamefont{Gehrmann et~al.}(2012)\citenamefont{Gehrmann, Luebbert,
  and Yang}}]{Gehrmann:2012ze}
\bibinfo{author}{\bibfnamefont{T.}~\bibnamefont{Gehrmann}},
  \bibinfo{author}{\bibfnamefont{T.}~\bibnamefont{Luebbert}}, \bibnamefont{and}
  \bibinfo{author}{\bibfnamefont{L.~L.} \bibnamefont{Yang}},
  \bibinfo{journal}{Phys. Rev. Lett.} \textbf{\bibinfo{volume}{109}},
  \bibinfo{pages}{242003} (\bibinfo{year}{2012}), \eprint{1209.0682}.

\bibitem[{\citenamefont{Gehrmann et~al.}(2014)\citenamefont{Gehrmann, Luebbert,
  and Yang}}]{Gehrmann:2014yya}
\bibinfo{author}{\bibfnamefont{T.}~\bibnamefont{Gehrmann}},
  \bibinfo{author}{\bibfnamefont{T.}~\bibnamefont{Luebbert}}, \bibnamefont{and}
  \bibinfo{author}{\bibfnamefont{L.~L.} \bibnamefont{Yang}},
  \bibinfo{journal}{JHEP} \textbf{\bibinfo{volume}{06}}, \bibinfo{pages}{155}
  (\bibinfo{year}{2014}), \eprint{1403.6451}.

\bibitem[{\citenamefont{Echevarria
  et~al.}(2014{\natexlab{b}})\citenamefont{Echevarria, Idilbi, and
  Scimemi}}]{Echevarria:2014rua}
\bibinfo{author}{\bibfnamefont{M.~G.} \bibnamefont{Echevarria}},
  \bibinfo{author}{\bibfnamefont{A.}~\bibnamefont{Idilbi}}, \bibnamefont{and}
  \bibinfo{author}{\bibfnamefont{I.}~\bibnamefont{Scimemi}},
  \bibinfo{journal}{Phys. Rev.} \textbf{\bibinfo{volume}{D90}},
  \bibinfo{pages}{014003} (\bibinfo{year}{2014}{\natexlab{b}}),
  \eprint{1402.0869}.

\bibitem[{\citenamefont{Echevarria
  et~al.}(2016{\natexlab{c}})\citenamefont{Echevarria, Scimemi, and
  Vladimirov}}]{Echevarria:2015usa}
\bibinfo{author}{\bibfnamefont{M.~G.} \bibnamefont{Echevarria}},
  \bibinfo{author}{\bibfnamefont{I.}~\bibnamefont{Scimemi}}, \bibnamefont{and}
  \bibinfo{author}{\bibfnamefont{A.}~\bibnamefont{Vladimirov}},
  \bibinfo{journal}{Phys. Rev.} \textbf{\bibinfo{volume}{D93}},
  \bibinfo{pages}{011502} (\bibinfo{year}{2016}{\natexlab{c}}),
  \bibinfo{note}{[Erratum: Phys. Rev.D94,no.9,099904(2016)]},
  \eprint{1509.06392}.

\bibitem[{\citenamefont{Collins and Metz}(2004)}]{Collins:2004nx}
\bibinfo{author}{\bibfnamefont{J.~C.} \bibnamefont{Collins}} \bibnamefont{and}
  \bibinfo{author}{\bibfnamefont{A.}~\bibnamefont{Metz}},
  \bibinfo{journal}{Phys. Rev. Lett.} \textbf{\bibinfo{volume}{93}},
  \bibinfo{pages}{252001} (\bibinfo{year}{2004}), \eprint{hep-ph/0408249}.

\bibitem[{\citenamefont{Gaunt}(2014)}]{Gaunt:2014ska}
\bibinfo{author}{\bibfnamefont{J.~R.} \bibnamefont{Gaunt}},
  \bibinfo{journal}{JHEP} \textbf{\bibinfo{volume}{07}}, \bibinfo{pages}{110}
  (\bibinfo{year}{2014}), \eprint{1405.2080}.

\bibitem[{\citenamefont{Diehl et~al.}(2016)\citenamefont{Diehl, Gaunt,
  Ostermeier, Plößl, and Schäfer}}]{Diehl:2015bca}
\bibinfo{author}{\bibfnamefont{M.}~\bibnamefont{Diehl}},
  \bibinfo{author}{\bibfnamefont{J.~R.} \bibnamefont{Gaunt}},
  \bibinfo{author}{\bibfnamefont{D.}~\bibnamefont{Ostermeier}},
  \bibinfo{author}{\bibfnamefont{P.}~\bibnamefont{Plößl}}, \bibnamefont{and}
  \bibinfo{author}{\bibfnamefont{A.}~\bibnamefont{Schäfer}},
  \bibinfo{journal}{JHEP} \textbf{\bibinfo{volume}{01}}, \bibinfo{pages}{076}
  (\bibinfo{year}{2016}), \eprint{1510.08696}.

\bibitem[{\citenamefont{Boer et~al.}(2017)\citenamefont{Boer, van Daal, Gaunt,
  Kasemets, and Mulders}}]{Boer:2017hqr}
\bibinfo{author}{\bibfnamefont{D.}~\bibnamefont{Boer}},
  \bibinfo{author}{\bibfnamefont{T.}~\bibnamefont{van Daal}},
  \bibinfo{author}{\bibfnamefont{J.~R.} \bibnamefont{Gaunt}},
  \bibinfo{author}{\bibfnamefont{T.}~\bibnamefont{Kasemets}}, \bibnamefont{and}
  \bibinfo{author}{\bibfnamefont{P.~J.} \bibnamefont{Mulders}},
  \bibinfo{journal}{SciPost Phys.} \textbf{\bibinfo{volume}{3}},
  \bibinfo{pages}{040} (\bibinfo{year}{2017}), \eprint{1709.04935}.

\bibitem[{\citenamefont{Boer et~al.}(2003)\citenamefont{Boer, Mulders, and
  Pijlman}}]{Boer:2003cm}
\bibinfo{author}{\bibfnamefont{D.}~\bibnamefont{Boer}},
  \bibinfo{author}{\bibfnamefont{P.~J.} \bibnamefont{Mulders}},
  \bibnamefont{and} \bibinfo{author}{\bibfnamefont{F.}~\bibnamefont{Pijlman}},
  \bibinfo{journal}{Nucl. Phys.} \textbf{\bibinfo{volume}{B667}},
  \bibinfo{pages}{201} (\bibinfo{year}{2003}), \eprint{hep-ph/0303034}.

\bibitem[{\citenamefont{Ji et~al.}(2006{\natexlab{a}})\citenamefont{Ji, Qiu,
  Vogelsang, and Yuan}}]{Ji:2006ub}
\bibinfo{author}{\bibfnamefont{X.}~\bibnamefont{Ji}},
  \bibinfo{author}{\bibfnamefont{J.-W.} \bibnamefont{Qiu}},
  \bibinfo{author}{\bibfnamefont{W.}~\bibnamefont{Vogelsang}},
  \bibnamefont{and} \bibinfo{author}{\bibfnamefont{F.}~\bibnamefont{Yuan}},
  \bibinfo{journal}{Phys. Rev. Lett.} \textbf{\bibinfo{volume}{97}},
  \bibinfo{pages}{082002} (\bibinfo{year}{2006}{\natexlab{a}}),
  \eprint{hep-ph/0602239}.

\bibitem[{\citenamefont{Ji et~al.}(2006{\natexlab{b}})\citenamefont{Ji, Qiu,
  Vogelsang, and Yuan}}]{Ji:2006vf}
\bibinfo{author}{\bibfnamefont{X.}~\bibnamefont{Ji}},
  \bibinfo{author}{\bibfnamefont{J.-w.} \bibnamefont{Qiu}},
  \bibinfo{author}{\bibfnamefont{W.}~\bibnamefont{Vogelsang}},
  \bibnamefont{and} \bibinfo{author}{\bibfnamefont{F.}~\bibnamefont{Yuan}},
  \bibinfo{journal}{Phys. Rev.} \textbf{\bibinfo{volume}{D73}},
  \bibinfo{pages}{094017} (\bibinfo{year}{2006}{\natexlab{b}}),
  \eprint{hep-ph/0604023}.

\bibitem[{\citenamefont{Koike et~al.}(2008)\citenamefont{Koike, Vogelsang, and
  Yuan}}]{Koike:2007dg}
\bibinfo{author}{\bibfnamefont{Y.}~\bibnamefont{Koike}},
  \bibinfo{author}{\bibfnamefont{W.}~\bibnamefont{Vogelsang}},
  \bibnamefont{and} \bibinfo{author}{\bibfnamefont{F.}~\bibnamefont{Yuan}},
  \bibinfo{journal}{Phys. Lett.} \textbf{\bibinfo{volume}{B659}},
  \bibinfo{pages}{878} (\bibinfo{year}{2008}), \eprint{0711.0636}.

\bibitem[{\citenamefont{Kang et~al.}(2011)\citenamefont{Kang, Xiao, and
  Yuan}}]{Kang:2011mr}
\bibinfo{author}{\bibfnamefont{Z.-B.} \bibnamefont{Kang}},
  \bibinfo{author}{\bibfnamefont{B.-W.} \bibnamefont{Xiao}}, \bibnamefont{and}
  \bibinfo{author}{\bibfnamefont{F.}~\bibnamefont{Yuan}},
  \bibinfo{journal}{Phys. Rev. Lett.} \textbf{\bibinfo{volume}{107}},
  \bibinfo{pages}{152002} (\bibinfo{year}{2011}), \eprint{1106.0266}.

\bibitem[{\citenamefont{Sun and Yuan}(2013)}]{Sun:2013hua}
\bibinfo{author}{\bibfnamefont{P.}~\bibnamefont{Sun}} \bibnamefont{and}
  \bibinfo{author}{\bibfnamefont{F.}~\bibnamefont{Yuan}},
  \bibinfo{journal}{Phys. Rev.} \textbf{\bibinfo{volume}{D88}},
  \bibinfo{pages}{114012} (\bibinfo{year}{2013}), \eprint{1308.5003}.

\bibitem[{\citenamefont{Dai et~al.}(2015)\citenamefont{Dai, Kang, Prokudin, and
  Vitev}}]{Dai:2014ala}
\bibinfo{author}{\bibfnamefont{L.-Y.} \bibnamefont{Dai}},
  \bibinfo{author}{\bibfnamefont{Z.-B.} \bibnamefont{Kang}},
  \bibinfo{author}{\bibfnamefont{A.}~\bibnamefont{Prokudin}}, \bibnamefont{and}
  \bibinfo{author}{\bibfnamefont{I.}~\bibnamefont{Vitev}},
  \bibinfo{journal}{Phys. Rev.} \textbf{\bibinfo{volume}{D92}},
  \bibinfo{pages}{114024} (\bibinfo{year}{2015}), \eprint{1409.5851}.

\bibitem[{\citenamefont{Scimemi and
  Vladimirov}(2018{\natexlab{c}})}]{Scimemi:2018mmi}
\bibinfo{author}{\bibfnamefont{I.}~\bibnamefont{Scimemi}} \bibnamefont{and}
  \bibinfo{author}{\bibfnamefont{A.}~\bibnamefont{Vladimirov}},
  \bibinfo{journal}{Eur. Phys. J.} \textbf{\bibinfo{volume}{C78}},
  \bibinfo{pages}{802} (\bibinfo{year}{2018}{\natexlab{c}}),
  \eprint{1804.08148}.

\bibitem[{\citenamefont{Bacchetta and Prokudin}(2013)}]{Bacchetta:2013pqa}
\bibinfo{author}{\bibfnamefont{A.}~\bibnamefont{Bacchetta}} \bibnamefont{and}
  \bibinfo{author}{\bibfnamefont{A.}~\bibnamefont{Prokudin}},
  \bibinfo{journal}{Nucl. Phys.} \textbf{\bibinfo{volume}{B875}},
  \bibinfo{pages}{536} (\bibinfo{year}{2013}), \eprint{1303.2129}.

\bibitem[{\citenamefont{Buffing
  et~al.}(2018{\natexlab{a}})\citenamefont{Buffing, Diehl, and
  Kasemets}}]{Buffing:2017mqm}
\bibinfo{author}{\bibfnamefont{M.~G.~A.} \bibnamefont{Buffing}},
  \bibinfo{author}{\bibfnamefont{M.}~\bibnamefont{Diehl}}, \bibnamefont{and}
  \bibinfo{author}{\bibfnamefont{T.}~\bibnamefont{Kasemets}},
  \bibinfo{journal}{JHEP} \textbf{\bibinfo{volume}{01}}, \bibinfo{pages}{044}
  (\bibinfo{year}{2018}{\natexlab{a}}), \eprint{1708.03528}.

\bibitem[{\citenamefont{Kanazawa et~al.}(2016)\citenamefont{Kanazawa, Koike,
  Metz, Pitonyak, and Schlegel}}]{Kanazawa:2015ajw}
\bibinfo{author}{\bibfnamefont{K.}~\bibnamefont{Kanazawa}},
  \bibinfo{author}{\bibfnamefont{Y.}~\bibnamefont{Koike}},
  \bibinfo{author}{\bibfnamefont{A.}~\bibnamefont{Metz}},
  \bibinfo{author}{\bibfnamefont{D.}~\bibnamefont{Pitonyak}}, \bibnamefont{and}
  \bibinfo{author}{\bibfnamefont{M.}~\bibnamefont{Schlegel}},
  \bibinfo{journal}{Phys. Rev.} \textbf{\bibinfo{volume}{D93}},
  \bibinfo{pages}{054024} (\bibinfo{year}{2016}), \eprint{1512.07233}.

\bibitem[{\citenamefont{Chai et~al.}(2018)\citenamefont{Chai, Chen, and
  Ma}}]{Chai:2018mwx}
\bibinfo{author}{\bibfnamefont{X.~P.} \bibnamefont{Chai}},
  \bibinfo{author}{\bibfnamefont{K.~B.} \bibnamefont{Chen}}, \bibnamefont{and}
  \bibinfo{author}{\bibfnamefont{J.~P.} \bibnamefont{Ma}}
  (\bibinfo{year}{2018}), \eprint{1808.10560}.

\bibitem[{\citenamefont{Scimemi et~al.}(2019)\citenamefont{Scimemi, Tarasov,
  and Vladimirov}}]{stv}
\bibinfo{author}{\bibfnamefont{I.}~\bibnamefont{Scimemi}},
  \bibinfo{author}{\bibfnamefont{A.}~\bibnamefont{Tarasov}}, \bibnamefont{and}
  \bibinfo{author}{\bibfnamefont{A.}~\bibnamefont{Vladimirov}}
  (\bibinfo{year}{2019}), \eprint{1901.04519}.

\bibitem[{\citenamefont{Becher et~al.}(2012)\citenamefont{Becher, Neubert, and
  Wilhelm}}]{Becher:2011xn}
\bibinfo{author}{\bibfnamefont{T.}~\bibnamefont{Becher}},
  \bibinfo{author}{\bibfnamefont{M.}~\bibnamefont{Neubert}}, \bibnamefont{and}
  \bibinfo{author}{\bibfnamefont{D.}~\bibnamefont{Wilhelm}},
  \bibinfo{journal}{JHEP} \textbf{\bibinfo{volume}{02}}, \bibinfo{pages}{124}
  (\bibinfo{year}{2012}), \eprint{1109.6027}.

\bibitem[{\citenamefont{Li and Zhu}(2017)}]{Li:2016ctv}
\bibinfo{author}{\bibfnamefont{Y.}~\bibnamefont{Li}} \bibnamefont{and}
  \bibinfo{author}{\bibfnamefont{H.~X.} \bibnamefont{Zhu}},
  \bibinfo{journal}{Phys. Rev. Lett.} \textbf{\bibinfo{volume}{118}},
  \bibinfo{pages}{022004} (\bibinfo{year}{2017}), \eprint{1604.01404}.

\bibitem[{\citenamefont{Vladimirov}(2017)}]{Vladimirov:2016dll}
\bibinfo{author}{\bibfnamefont{A.~A.} \bibnamefont{Vladimirov}},
  \bibinfo{journal}{Phys. Rev. Lett.} \textbf{\bibinfo{volume}{118}},
  \bibinfo{pages}{062001} (\bibinfo{year}{2017}), \eprint{1610.05791}.

\bibitem[{\citenamefont{Moch and Vermaseren}(2000)}]{Moch:1999eb}
\bibinfo{author}{\bibfnamefont{S.}~\bibnamefont{Moch}} \bibnamefont{and}
  \bibinfo{author}{\bibfnamefont{J.~A.~M.} \bibnamefont{Vermaseren}},
  \bibinfo{journal}{Nucl. Phys.} \textbf{\bibinfo{volume}{B573}},
  \bibinfo{pages}{853} (\bibinfo{year}{2000}), \eprint{hep-ph/9912355}.

\bibitem[{\citenamefont{Mitov and Moch}(2006)}]{Mitov:2006wy}
\bibinfo{author}{\bibfnamefont{A.}~\bibnamefont{Mitov}} \bibnamefont{and}
  \bibinfo{author}{\bibfnamefont{S.-O.} \bibnamefont{Moch}},
  \bibinfo{journal}{Nucl. Phys.} \textbf{\bibinfo{volume}{B751}},
  \bibinfo{pages}{18} (\bibinfo{year}{2006}), \eprint{hep-ph/0604160}.

\bibitem[{\citenamefont{Drell and Yan}(1970)}]{Drell:1970wh}
\bibinfo{author}{\bibfnamefont{S.~D.} \bibnamefont{Drell}} \bibnamefont{and}
  \bibinfo{author}{\bibfnamefont{T.-M.} \bibnamefont{Yan}},
  \bibinfo{journal}{Phys. Rev. Lett.} \textbf{\bibinfo{volume}{25}},
  \bibinfo{pages}{316} (\bibinfo{year}{1970}), \bibinfo{note}{[Erratum: Phys.
  Rev. Lett.25,902(1970)]}.

\bibitem[{\citenamefont{Altarelli et~al.}(1978)\citenamefont{Altarelli, Ellis,
  and Martinelli}}]{Altarelli:1978id}
\bibinfo{author}{\bibfnamefont{G.}~\bibnamefont{Altarelli}},
  \bibinfo{author}{\bibfnamefont{R.~K.} \bibnamefont{Ellis}}, \bibnamefont{and}
  \bibinfo{author}{\bibfnamefont{G.}~\bibnamefont{Martinelli}},
  \bibinfo{journal}{Nucl. Phys.} \textbf{\bibinfo{volume}{B143}},
  \bibinfo{pages}{521} (\bibinfo{year}{1978}), \bibinfo{note}{[Erratum: Nucl.
  Phys.B146,544(1978)]}.

\bibitem[{\citenamefont{Tangerman and Mulders}(1995)}]{Tangerman:1994eh}
\bibinfo{author}{\bibfnamefont{R.~D.} \bibnamefont{Tangerman}}
  \bibnamefont{and} \bibinfo{author}{\bibfnamefont{P.~J.}
  \bibnamefont{Mulders}}, \bibinfo{journal}{Phys. Rev.}
  \textbf{\bibinfo{volume}{D51}}, \bibinfo{pages}{3357} (\bibinfo{year}{1995}),
  \eprint{hep-ph/9403227}.

\bibitem[{\citenamefont{Kramer and Lampe}(1987)}]{Kramer:1986sg}
\bibinfo{author}{\bibfnamefont{G.}~\bibnamefont{Kramer}} \bibnamefont{and}
  \bibinfo{author}{\bibfnamefont{B.}~\bibnamefont{Lampe}}, \bibinfo{journal}{Z.
  Phys.} \textbf{\bibinfo{volume}{C34}}, \bibinfo{pages}{497}
  (\bibinfo{year}{1987}), \bibinfo{note}{[Erratum: Z. Phys.C42,504(1989)]}.

\bibitem[{\citenamefont{Matsuura et~al.}(1989)\citenamefont{Matsuura, van~der
  Marck, and van Neerven}}]{Matsuura:1988sm}
\bibinfo{author}{\bibfnamefont{T.}~\bibnamefont{Matsuura}},
  \bibinfo{author}{\bibfnamefont{S.~C.} \bibnamefont{van~der Marck}},
  \bibnamefont{and} \bibinfo{author}{\bibfnamefont{W.~L.} \bibnamefont{van
  Neerven}}, \bibinfo{journal}{Nucl. Phys.} \textbf{\bibinfo{volume}{B319}},
  \bibinfo{pages}{570} (\bibinfo{year}{1989}).

\bibitem[{\citenamefont{Idilbi et~al.}(2006)\citenamefont{Idilbi, Ji, and
  Yuan}}]{Idilbi:2006dg}
\bibinfo{author}{\bibfnamefont{A.}~\bibnamefont{Idilbi}},
  \bibinfo{author}{\bibfnamefont{X.-d.} \bibnamefont{Ji}}, \bibnamefont{and}
  \bibinfo{author}{\bibfnamefont{F.}~\bibnamefont{Yuan}},
  \bibinfo{journal}{Nucl. Phys.} \textbf{\bibinfo{volume}{B753}},
  \bibinfo{pages}{42} (\bibinfo{year}{2006}), \eprint{hep-ph/0605068}.

\bibitem[{\citenamefont{Balitsky and Tarasov}(2018)}]{Balitsky:2017gis}
\bibinfo{author}{\bibfnamefont{I.}~\bibnamefont{Balitsky}} \bibnamefont{and}
  \bibinfo{author}{\bibfnamefont{A.}~\bibnamefont{Tarasov}},
  \bibinfo{journal}{JHEP} \textbf{\bibinfo{volume}{05}}, \bibinfo{pages}{150}
  (\bibinfo{year}{2018}), \eprint{1712.09389}.

\bibitem[{\citenamefont{Ebert et~al.}(2018{\natexlab{b}})\citenamefont{Ebert,
  Moult, Stewart, Tackmann, Vita, and Zhu}}]{Ebert:2018gsn}
\bibinfo{author}{\bibfnamefont{M.~A.} \bibnamefont{Ebert}},
  \bibinfo{author}{\bibfnamefont{I.}~\bibnamefont{Moult}},
  \bibinfo{author}{\bibfnamefont{I.~W.} \bibnamefont{Stewart}},
  \bibinfo{author}{\bibfnamefont{F.~J.} \bibnamefont{Tackmann}},
  \bibinfo{author}{\bibfnamefont{G.}~\bibnamefont{Vita}}, \bibnamefont{and}
  \bibinfo{author}{\bibfnamefont{H.~X.} \bibnamefont{Zhu}}
  (\bibinfo{year}{2018}{\natexlab{b}}), \eprint{1812.08189}.

\bibitem[{\citenamefont{Collins et~al.}(2016)\citenamefont{Collins, Gamberg,
  Prokudin, Rogers, Sato, and Wang}}]{Collins:2016hqq}
\bibinfo{author}{\bibfnamefont{J.}~\bibnamefont{Collins}},
  \bibinfo{author}{\bibfnamefont{L.}~\bibnamefont{Gamberg}},
  \bibinfo{author}{\bibfnamefont{A.}~\bibnamefont{Prokudin}},
  \bibinfo{author}{\bibfnamefont{T.~C.} \bibnamefont{Rogers}},
  \bibinfo{author}{\bibfnamefont{N.}~\bibnamefont{Sato}}, \bibnamefont{and}
  \bibinfo{author}{\bibfnamefont{B.}~\bibnamefont{Wang}},
  \bibinfo{journal}{Phys. Rev.} \textbf{\bibinfo{volume}{D94}},
  \bibinfo{pages}{034014} (\bibinfo{year}{2016}), \eprint{1605.00671}.

\bibitem[{\citenamefont{Gamberg et~al.}(2018)\citenamefont{Gamberg, Metz,
  Pitonyak, and Prokudin}}]{Gamberg:2017jha}
\bibinfo{author}{\bibfnamefont{L.}~\bibnamefont{Gamberg}},
  \bibinfo{author}{\bibfnamefont{A.}~\bibnamefont{Metz}},
  \bibinfo{author}{\bibfnamefont{D.}~\bibnamefont{Pitonyak}}, \bibnamefont{and}
  \bibinfo{author}{\bibfnamefont{A.}~\bibnamefont{Prokudin}},
  \bibinfo{journal}{Phys. Lett.} \textbf{\bibinfo{volume}{B781}},
  \bibinfo{pages}{443} (\bibinfo{year}{2018}), \eprint{1712.08116}.

\bibitem[{\citenamefont{Davies et~al.}(1985)\citenamefont{Davies, Webber, and
  Stirling}}]{Davies:1984sp}
\bibinfo{author}{\bibfnamefont{C.~T.~H.} \bibnamefont{Davies}},
  \bibinfo{author}{\bibfnamefont{B.~R.} \bibnamefont{Webber}},
  \bibnamefont{and} \bibinfo{author}{\bibfnamefont{W.~J.}
  \bibnamefont{Stirling}}, \bibinfo{journal}{Nucl. Phys.}
  \textbf{\bibinfo{volume}{B256}}, \bibinfo{pages}{413} (\bibinfo{year}{1985}),
  \bibinfo{note}{[1,I.95(1984)]}.

\bibitem[{\citenamefont{Ellis and Veseli}(1998)}]{Ellis:1997ii}
\bibinfo{author}{\bibfnamefont{R.~K.} \bibnamefont{Ellis}} \bibnamefont{and}
  \bibinfo{author}{\bibfnamefont{S.}~\bibnamefont{Veseli}},
  \bibinfo{journal}{Nucl. Phys.} \textbf{\bibinfo{volume}{B511}},
  \bibinfo{pages}{649} (\bibinfo{year}{1998}), \eprint{hep-ph/9706526}.

\bibitem[{\citenamefont{Korchemsky and Sterman}(1995)}]{Korchemsky:1994is}
\bibinfo{author}{\bibfnamefont{G.~P.} \bibnamefont{Korchemsky}}
  \bibnamefont{and} \bibinfo{author}{\bibfnamefont{G.~F.}
  \bibnamefont{Sterman}}, \bibinfo{journal}{Nucl. Phys.}
  \textbf{\bibinfo{volume}{B437}}, \bibinfo{pages}{415} (\bibinfo{year}{1995}),
  \eprint{hep-ph/9411211}.

\bibitem[{\citenamefont{Collins and Soper}(1981)}]{Collins:1981uk}
\bibinfo{author}{\bibfnamefont{J.~C.} \bibnamefont{Collins}} \bibnamefont{and}
  \bibinfo{author}{\bibfnamefont{D.~E.} \bibnamefont{Soper}},
  \bibinfo{journal}{Nucl. Phys.} \textbf{\bibinfo{volume}{B193}},
  \bibinfo{pages}{381} (\bibinfo{year}{1981}), \bibinfo{note}{[Erratum: Nucl.
  Phys.B213,545(1983)]}.

\bibitem[{\citenamefont{Li et~al.}(2016)\citenamefont{Li, Neill, and
  Zhu}}]{Li:2016axz}
\bibinfo{author}{\bibfnamefont{Y.}~\bibnamefont{Li}},
  \bibinfo{author}{\bibfnamefont{D.}~\bibnamefont{Neill}}, \bibnamefont{and}
  \bibinfo{author}{\bibfnamefont{H.~X.} \bibnamefont{Zhu}},
  \bibinfo{journal}{Submitted to: Phys. Rev. D}  (\bibinfo{year}{2016}),
  \eprint{1604.00392}.

\bibitem[{\citenamefont{Bacchetta et~al.}(2017)\citenamefont{Bacchetta,
  Delcarro, Pisano, Radici, and Signori}}]{Bacchetta:2017gcc}
\bibinfo{author}{\bibfnamefont{A.}~\bibnamefont{Bacchetta}},
  \bibinfo{author}{\bibfnamefont{F.}~\bibnamefont{Delcarro}},
  \bibinfo{author}{\bibfnamefont{C.}~\bibnamefont{Pisano}},
  \bibinfo{author}{\bibfnamefont{M.}~\bibnamefont{Radici}}, \bibnamefont{and}
  \bibinfo{author}{\bibfnamefont{A.}~\bibnamefont{Signori}}
  (\bibinfo{year}{2017}), \eprint{1703.10157}.

\bibitem[{\citenamefont{Becher and Bell}(2014)}]{Becher:2013iya}
\bibinfo{author}{\bibfnamefont{T.}~\bibnamefont{Becher}} \bibnamefont{and}
  \bibinfo{author}{\bibfnamefont{G.}~\bibnamefont{Bell}},
  \bibinfo{journal}{Phys. Rev. Lett.} \textbf{\bibinfo{volume}{112}},
  \bibinfo{pages}{182002} (\bibinfo{year}{2014}), \eprint{1312.5327}.

\bibitem[{\citenamefont{Aad et~al.}(2016)}]{Aad:2015auj}
\bibinfo{author}{\bibfnamefont{G.}~\bibnamefont{Aad}} \bibnamefont{et~al.}
  (\bibinfo{collaboration}{ATLAS}), \bibinfo{journal}{Eur. Phys. J.}
  \textbf{\bibinfo{volume}{C76}}, \bibinfo{pages}{291} (\bibinfo{year}{2016}),
  \eprint{1512.02192}.

\bibitem[{\citenamefont{Nadolsky et~al.}(2001)\citenamefont{Nadolsky, Stump,
  and Yuan}}]{Nadolsky:2000ky}
\bibinfo{author}{\bibfnamefont{P.~M.} \bibnamefont{Nadolsky}},
  \bibinfo{author}{\bibfnamefont{D.~R.} \bibnamefont{Stump}}, \bibnamefont{and}
  \bibinfo{author}{\bibfnamefont{C.~P.} \bibnamefont{Yuan}},
  \bibinfo{journal}{Phys. Rev.} \textbf{\bibinfo{volume}{D64}},
  \bibinfo{pages}{114011} (\bibinfo{year}{2001}), \eprint{hep-ph/0012261}.

\bibitem[{\citenamefont{Bozzi et~al.}(2011)\citenamefont{Bozzi, Catani,
  Ferrera, de~Florian, and Grazzini}}]{Bozzi:2010xn}
\bibinfo{author}{\bibfnamefont{G.}~\bibnamefont{Bozzi}},
  \bibinfo{author}{\bibfnamefont{S.}~\bibnamefont{Catani}},
  \bibinfo{author}{\bibfnamefont{G.}~\bibnamefont{Ferrera}},
  \bibinfo{author}{\bibfnamefont{D.}~\bibnamefont{de~Florian}},
  \bibnamefont{and} \bibinfo{author}{\bibfnamefont{M.}~\bibnamefont{Grazzini}},
  \bibinfo{journal}{Phys. Lett.} \textbf{\bibinfo{volume}{B696}},
  \bibinfo{pages}{207} (\bibinfo{year}{2011}), \eprint{1007.2351}.

\bibitem[{\citenamefont{Kang et~al.}(2016)\citenamefont{Kang, Qiu, Wang, and
  Xing}}]{Kang:2016ron}
\bibinfo{author}{\bibfnamefont{Z.-B.} \bibnamefont{Kang}},
  \bibinfo{author}{\bibfnamefont{J.-W.} \bibnamefont{Qiu}},
  \bibinfo{author}{\bibfnamefont{X.-N.} \bibnamefont{Wang}}, \bibnamefont{and}
  \bibinfo{author}{\bibfnamefont{H.}~\bibnamefont{Xing}},
  \bibinfo{journal}{Phys. Rev.} \textbf{\bibinfo{volume}{D94}},
  \bibinfo{pages}{074038} (\bibinfo{year}{2016}), \eprint{1605.07175}.

\bibitem[{\citenamefont{Kang et~al.}(2017{\natexlab{a}})\citenamefont{Kang,
  Liu, Ringer, and Xing}}]{Kang:2017glf}
\bibinfo{author}{\bibfnamefont{Z.-B.} \bibnamefont{Kang}},
  \bibinfo{author}{\bibfnamefont{X.}~\bibnamefont{Liu}},
  \bibinfo{author}{\bibfnamefont{F.}~\bibnamefont{Ringer}}, \bibnamefont{and}
  \bibinfo{author}{\bibfnamefont{H.}~\bibnamefont{Xing}},
  \bibinfo{journal}{JHEP} \textbf{\bibinfo{volume}{11}}, \bibinfo{pages}{068}
  (\bibinfo{year}{2017}{\natexlab{a}}), \eprint{1705.08443}.

\bibitem[{\citenamefont{Kang et~al.}(2017{\natexlab{b}})\citenamefont{Kang,
  Prokudin, Ringer, and Yuan}}]{Kang:2017btw}
\bibinfo{author}{\bibfnamefont{Z.-B.} \bibnamefont{Kang}},
  \bibinfo{author}{\bibfnamefont{A.}~\bibnamefont{Prokudin}},
  \bibinfo{author}{\bibfnamefont{F.}~\bibnamefont{Ringer}}, \bibnamefont{and}
  \bibinfo{author}{\bibfnamefont{F.}~\bibnamefont{Yuan}},
  \bibinfo{journal}{Phys. Lett.} \textbf{\bibinfo{volume}{B774}},
  \bibinfo{pages}{635} (\bibinfo{year}{2017}{\natexlab{b}}),
  \eprint{1707.00913}.

\bibitem[{\citenamefont{Gutierrez-Reyes
  et~al.}(2018{\natexlab{b}})\citenamefont{Gutierrez-Reyes, Scimemi, Waalewijn,
  and Zoppi}}]{Gutierrez-Reyes:2018qez}
\bibinfo{author}{\bibfnamefont{D.}~\bibnamefont{Gutierrez-Reyes}},
  \bibinfo{author}{\bibfnamefont{I.}~\bibnamefont{Scimemi}},
  \bibinfo{author}{\bibfnamefont{W.~J.} \bibnamefont{Waalewijn}},
  \bibnamefont{and} \bibinfo{author}{\bibfnamefont{L.}~\bibnamefont{Zoppi}},
  \bibinfo{journal}{Phys. Rev. Lett.} \textbf{\bibinfo{volume}{121}},
  \bibinfo{pages}{162001} (\bibinfo{year}{2018}{\natexlab{b}}),
  \eprint{1807.07573}.

\bibitem[{\citenamefont{Liu et~al.}(2018)\citenamefont{Liu, Ringer, Vogelsang,
  and Yuan}}]{Liu:2018trl}
\bibinfo{author}{\bibfnamefont{X.}~\bibnamefont{Liu}},
  \bibinfo{author}{\bibfnamefont{F.}~\bibnamefont{Ringer}},
  \bibinfo{author}{\bibfnamefont{W.}~\bibnamefont{Vogelsang}},
  \bibnamefont{and} \bibinfo{author}{\bibfnamefont{F.}~\bibnamefont{Yuan}}
  (\bibinfo{year}{2018}), \eprint{1812.08077}.

\bibitem[{\citenamefont{Buffing
  et~al.}(2018{\natexlab{b}})\citenamefont{Buffing, Kang, Lee, and
  Liu}}]{Buffing:2018ggv}
\bibinfo{author}{\bibfnamefont{M.~G.~A.} \bibnamefont{Buffing}},
  \bibinfo{author}{\bibfnamefont{Z.-B.} \bibnamefont{Kang}},
  \bibinfo{author}{\bibfnamefont{K.}~\bibnamefont{Lee}}, \bibnamefont{and}
  \bibinfo{author}{\bibfnamefont{X.}~\bibnamefont{Liu}}
  (\bibinfo{year}{2018}{\natexlab{b}}), \eprint{1812.07549}.

\end{thebibliography}
\end{document}